\documentclass{aa}

\usepackage{graphicx}
\usepackage{txfonts}
\usepackage{gensymb,amssymb}
\usepackage{xcolor}
\usepackage{lscape}
\usepackage[version=3]{mhchem}
\usepackage{caption,subcaption}
\usepackage{multirow}
\usepackage{soul}

\usepackage[colorlinks = true,
            linkcolor = blue,
            urlcolor  = blue,
            citecolor = blue,
            anchorcolor = blue]{hyperref}

\begin{document}

\title{The asymmetric carbon-rich chemistry of the planet-forming disk of HD~142527 triggered by late infall}

\author{Milou Temmink\orcid{0000-0002-7935-7445}\inst{1} \and
        Ewine F. van Dishoeck\orcid{0000-0001-7591-1907}\inst{1,2} \and
        Alice S. Booth\orcid{0000-0003-2014-2121}\inst{3}\thanks{Clay Postdoctoral Fellow}  \and 
        Nienke van der Marel\orcid{0000-0003-2458-9756}\inst{1} \and
        Myriam Benisty\orcid{0000-0002-7695-7605}\inst{4} \and
        Michiel R. Hogerheijde\orcid{0000-0001-5217-537X}\inst{1,5}}
\institute{Leiden Observatory, Leiden University, PO Box 9513, 2300 RA Leiden, the Netherlands \\
          \email{temmink@strw.leidenuniv.nl} \and
          Max-Planck-Institut f\"ur Extraterrestrische Physik, Giessenbachstraße 1, D-85748 Garching, Germany \and
          Center for Astrophysics - Harvard \& Smithsonian, 60 Garden St., Cambridge, MA 02138, USA \and
          Max-Planck-Institut f\"{u}r Astronomie (MPIA), K\"{o}nigstuhl 17, 69117 Heidelberg, Germany \and
          Anton Pannekoek Institute, University of Amsterdam, Science Park NL-904 1098XH, the Netherlands}
\date{Received 6 March, 2026; accepted 28 May, 2026}

%% ---

\abstract
{The planet-forming disk of HD 142527 is known for its azimuthally asymmetric dust trap, shadows, and spiral arms. In this work, we use new observations of the Atacama Large Millimeter/submillimeter Array to investigate the molecular composition and to determine the ongoing chemical processes and the origin of its asymmetric molecular emission, and to infer possible effects of dust continuum obscuration. The observations cover a wide variety of molecular species over a large frequency range, enlarging the known molecular inventory of this system. Strikingly, the emission of \ce{H_2CO}, \ce{CN}, and \ce{C_2H} is dominated by spiral-like features peaking in the southern region of the disk, opposite to the large dust trap, while no relation is found between the observed asymmetries and the shadows seen in the scattered light due to the misaligned inner disk. We attribute these features to low-density, late infalling, atomic carbon-rich material that locally enhances the C/O-ratio and, subsequently, facilitates the gas-phase formation of these species. Azimuthal offsets between the peak emission of \ce{H_2CO} and that of \ce{CN} and \ce{C_2H} are possibly due to a delay of a few hundred years in the gas-phase formation of \ce{H_2CO}. As opposed to the emission of \ce{H_2CO}, \ce{CN}, and \ce{C_2H}, the emission of \ce{C^{17}O} and the \ce{HCO^+} $J$=1-0 transition is aligned with the large dust trap, likely due to an azimuthal enhancement in the surface density. Differences between the two observed \ce{C^{17}O} transitions may be due to dust obscuration effects. The latter effect is not expected to affect molecular emission at 3 millimetres, given the lower optical depth of the dust trap. The four observed transitions of \ce{CS} display different azimuthal extents and strengths, with the lines with lower upper level energies appearing more ring-like. An analysis of the \ce{^{13}CO} brightness temperature yields no significant temperature variations across the disk's azimuth. Therefore, we propose that the observed \ce{CS} transitions trace two different reservoirs: a cold reservoir that resides on a Keplerian orbit and a second, hotter reservoir of \ce{CS} that is facilitated by the infalling material and resides in a higher atmospheric layer of the disk. A single weak transition of \ce{SO} is observed, which may be explained by weak shocks induced by the spirals observed in the scattered light that liberate sulphur. Future higher-resolution, multi-line observations of species such as \ce{H_2CO}, \ce{CS}, \ce{CN}, and \ce{C_2H} are needed to investigate the role and importance of late infalling material in setting the chemical composition of planet-forming disks.}


%% ---

\maketitle

\section{Introduction} \label{sec:Intro}
\begin{figure*}[ht!]
    \centering
    \includegraphics[width=\textwidth]{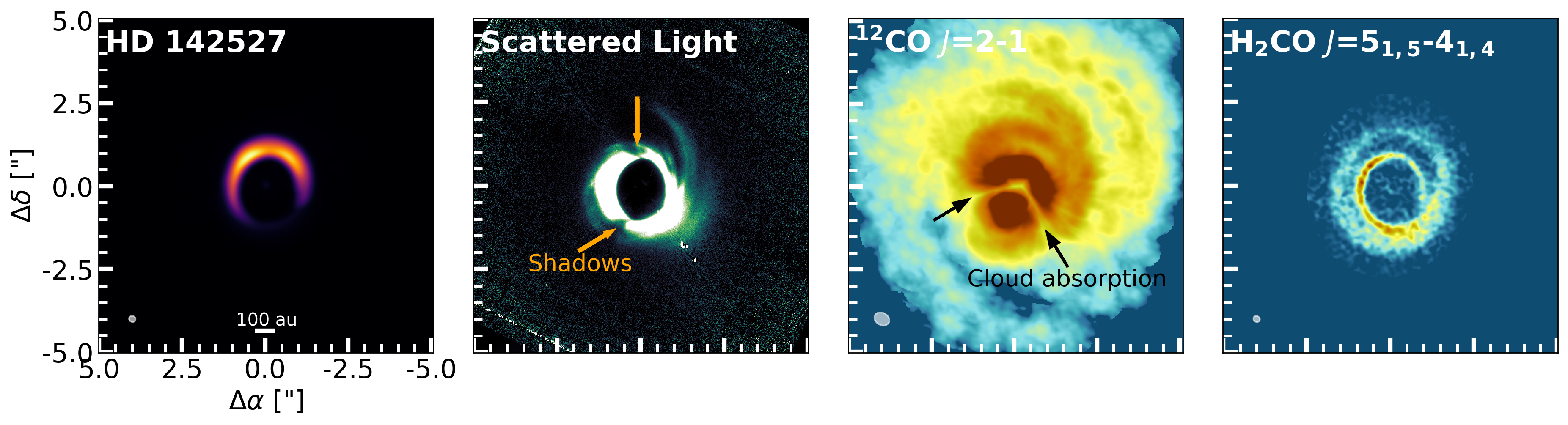}
    \caption{Continuum emission (left), scattered light (left central; $H$-band with SPHERE/IRDIS, \citealt{HunzikerEA17}), peak intensity map of the \ce{^{12}CO} $J$=2-1 transition (right central; see also \citealt{ChristiaensEA14,GargEA21}), and integrated intensity map of the \ce{H_2CO} $J$=5$_{1,5}$-4$_{1,4}$ transition (right) within the disk of HD~142527. The \ce{^{12}CO} map has been imaged using a square-root scaling scheme for the colour map. Additionally, the scattered light image has been scaled using the radial-distance squared, accounting for the drop off in stellar flux. We note that a few imaging artefacts are visible in the scattered light image. The arrows point, respectively, to the shadows observed in the scattered light and the dark emission lanes due to cloud absorption in the \ce{^{12}CO} image.}
    \label{fig:IntroOverview}
\end{figure*}
As giant planets accrete their atmospheres from the gaseous reservoirs of planet-forming disks, understanding the molecular composition and, in particular, the processes that set this composition is of great importance for our understanding of planet formation. The Atacama Large Millimeter/submillimeter Array (ALMA) yields unique insights to study the chemistry in the cold outer disks with unprecedented sensitivity and resolution. Recent programs have unveiled the chemical inventories in both individual disks (see, for example, \citealt{QiEA13,ObergEA15,BerginEA16,WalshEA16,CleevesEA18,KastnerEA18,LoomisEA18,SemenovEA18,FacchiniEA21,BoothEA24a,BoothEA24b,BoothEA25,RampinelliEA24}) and larger samples (see, for example, \citealt{LeGalEA19,ObergEA21,BoothEA26}). However, the dominant processes that set the observable chemistry are still far from being fully understood. \\
\indent One disk that can improve our understanding of the dominant processes in planet-forming disks is that around the young star HD~142527. This system is located at a distance of 159.26 pc \citep{GAIA23} and consists of a young F6 star \citep{FairlambEA15}, an M-dwarf companion \citep{BillerEA12,LacourEA16,BalmerEA22,StolkerEA24}, and a massive planet-forming disk ($M_\mathrm{gas}$=(1.6$\pm$0.6)$\times$10$^{-2}$ M$_\odot$; \citealt{TemminkEA23}). The mass of the host star is, however, uncertain, as recent papers yield a range of stellar masses from 1.69 M$_\odot$ to 2.40 M$_\odot$ \citep{FukagawaEA06,VerhoeffEA11,ArunEA19,FvdM20}. In this work, we assume the recent value from \citet{VioqueEA25} ($M_*\sim$2.24 M$_\odot$) as their astrometry results also yield a mass of the M-dwarf companion ($\sim$158 M$_\mathrm{Jup}$ or $\sim$0.15 M$_\odot$) that is dependent on the mass of the host star. The importance of this moderately inclined ($i\sim$28\degree) disk stems from the various key features it hosts (see Fig. \ref{fig:IntroOverview}): an asymmetric dust trap \citep{FujiwaraEA06,Ohashi08,CasassusEA13}, spiral arms in both the scattered light and the \ce{^{12}CO} molecular emission \citep{AvenhausEA14,ChristiaensEA14,GargEA21}, and shadows due to a misaligned inner disk \citep{MarinoEA15,BohnEA22}. The hydrodynamical models of \citet{PriceEA18} show that all these features can be attributed to the M-dwarf companion. A thorough investigation of the molecular emission may reveal unique insights into the importance of these features in setting the observable chemistry. \\
\indent Disks with asymmetries, both in the dust and in the gas, are unique laboratories to study the role of the dust in setting the observable chemistry. This has become most apparent in the case of Oph-IRS~48 (A0-type star), the most asymmetric system known to date \citep{vdMarelEA13}, where the molecular emission, except \ce{CO}, is approximately co-spatial with the dust continuum emission located at $\sim$60~au \citep{vdMarelEA21,BoothEA21,BrunkenEA22,LeemkerEA23,BoothEA24b}. As the emission of both simple and more complex species is co-spatial with the continuum, the dust trap has been proposed to be an ice trap, where radial and vertical transport leads to the sublimation of the icy mantles coating the dust grains. In addition, a more recent work also discusses the role of photodissociation and the subsequent gas-phase reactions involving the dissociation products in setting the observable molecular composition \citep{TemminkEA25}. \\
\indent Aside from dust traps, asymmetries may also be related to decoupled dust dynamics surrounding a binary-carved cavity \citep{PriceEA18}, azimuthal variations in the gas density, the temperature, and the incident ultraviolet (UV) radiation. A change in the temperature and UV radiation follows most likely from shadowing effects, which are often observed in the scattered light (see \citealt{BenistyEA23} for a recent review) and are attributed to misaligned inner disks and/or warps \citep{MarinoEA15,BohnEA22}. Recent modelling work has shown that an azimuthally varying temperature structure yields asymmetric column densities \citep{YoungEA21}. Extended structures, such as spirals and streamers, may also influence the observable chemistry and locally enhance the emission, for example, through shocks. Recently, two studies suggest a potential connection between observed \ce{SO} emission and the spiral arms in the disks of AB~Aur \citep{SpeedieEA25} and MWC~758 \citep{ZagariaEA25}. Additionally, \citet{IleeEA17} studied the influence of gravitational instabilities on the chemistry in protoplanetary fragments, finding that certain molecular species (e.g., \ce{H_2O}, \ce{H_2S}, \ce{SO}) are abundant in the fragments, while others (such as \ce{CO}, \ce{CH_4}, \ce{CN}, \ce{CS}, and \ce{H_2CO}) are more abundant in spiral shocks. Streamers (or late infalling material), on the other hand, may replenish the disk with fresh material that locally alters the elemental and molecular abundances, leading to molecular asymmetries. \\
\indent In this work, we use new ALMA observations to study the molecular emission of HD~142527 and to expand upon the known molecular inventory \citep{CasassusEA13,vdPlasEA14,TemminkEA23}. Using these observations, we propose various scenarios that may explain the observed molecular morphologies and try to answer the question of what sets the asymmetric molecular emission in the disk of HD~142527. \\
\indent This paper is structured as follows: the observations and self-calibration process are described in Section \ref{sec:Obs}, while we report our (weak) detected molecular species and non-detections in Section \ref{sec:Molecules}. Section \ref{sec:Disc} contains the analysis and discussion of the observed molecular species, including various scenarios to explain the origins of the emission. Finally, we summarise our findings and conclusions in Section \ref{sec:Summary}.

\section{Observations and self-calibration} \label{sec:Obs}
\begin{figure*}[ht!]
    \centering
    \includegraphics[width=\textwidth]{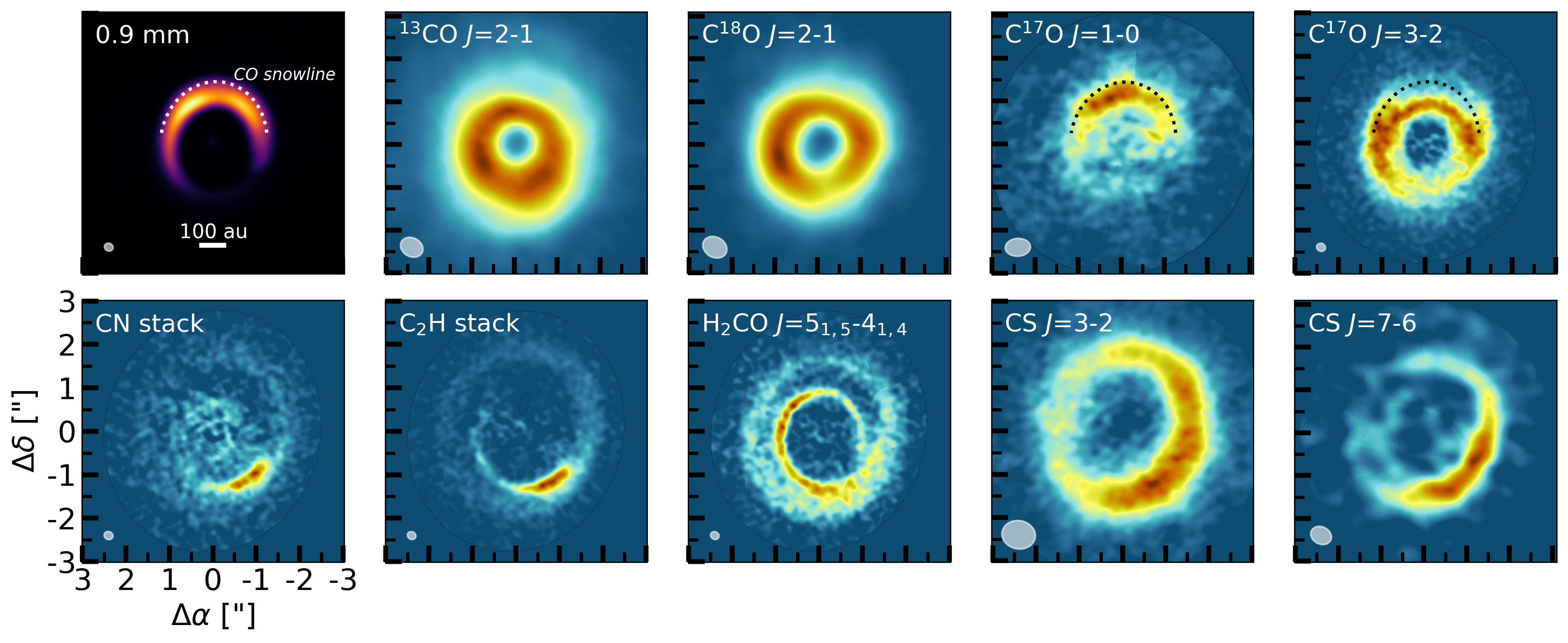}
    \caption{Integrated intensity maps of the dust continuum at 1.3~mm and key molecular species detected in the disk of HD~142527. The dotted arcs in the images of the continuum and \ce{C^{17}O} represent the approximate location of the \ce{CO} snowline based on the RADMC-3D model of \citet{TemminkEA23}. To increase the $S/N$-ratio of the respective integrated intensity maps, the images of the \ce{C^{17}O} $J$=1-0, \ce{C_2H} and \ce{CN} were created by stacking the detected transitions (two, four, and three transitions, respectively). Additionally, the displayed images of \ce{C^{17}O} $J$=1-0 and \ce{CS} $J$=2-1 transitions were created using a robust value of 2.0. The resolving beams are shown in the lower-left corner of each image.}
    \label{fig:KeyGallery}
\end{figure*}
The disk of HD~142527 was observed in two separate Cycle 10 and 11 ALMA programs, 2023.1.00628.S and 2024.1.00446.S (PI: M. Temmink). The first program covers two spectral settings in Band~3 and one spectral setting in Band~4, while the second program consists of one spectral setting in Band~7. Both programs have been taken in the C3 and C6 configurations, allowing for high spatial resolution observations of $\sim$0.15-0.20". The spectral windows in Band 3 have resolutions of 70.56~kHz and 564.45~kHz, resulting in velocity resolutions of, respectively, $\sim$0.22~km~s$^{-1}$ (at 97.98~GHz) and $\sim$1.96~km~s$^{-1}$ (at 86.43~GHz). Those in Band 4 have spectral resolutions of 141.113~kHz ($\sim$0.29~km~s$^{-1}$ at 146.95~GHz) and 564.453~kHz ($\sim$1.17~km~s$^{-1}$ at 144.52~GHz). Finally, the Band 7 spectral windows have resolutions of 141.113~kHz and 1.129~MHz, yielding velocity resolutions of $\sim$0.12~km~s$^{-1}$ at 349.36~GHz and $\sim$0.96~km~s$^{-1}$ at 351.55~GHz, respectively. The observations were reduced with the provided pipeline scripts using the specified Common Astronomy Software Applications (\textsc{CASA}; \citealt{McMullinEA07,CASA}). Self-calibration and imaging have, on the other hand, been carried out with \textsc{CASA} versions 6.4.1.12 and 6.5.4. Further details on the observations can be found in Table \ref{tab:ObsDetails}. Furthermore, we make use of the following archival programs: 2011.1.00318.S (PI: M. Fukagawa), 2011.0.00465.S (PI: S. Casassus), 2012.1.00631.S (PI: M. Fukagawa), 2013.1.00305.S (PI: S. Casassus), 2015.1.00805.S (PI: S. Casassus), and 2015.1.01137.S  (PI: T. Tsukagoshi). \\
\indent To increase the signal-to-noise ratio ($S/N$-ratio) of the observations, we employ similar self-calibration techniques as used in various successful ALMA programs \citep{AndrewsEA18,CzekalaEA21,LoomisEA25,LeemkerEA25}. To summarise, we performed the self-calibration routine on the short-baseline observations before combining the short- and long-baseline observations and performing the routine on the combined dataset. If an observation consists of multiple execution blocks, a single round of phase-only self-calibration was performed on each execution block before the execution blocks were aligned (if needed) and combined. The self-calibration routine consists of multiple rounds of phase-only self-calibration, using solution intervals given in the weblog of the observations, followed by a single round of phase-amplitude self-calibration (\textsc{solint}=`inf'). During the rounds of phase-only self-calibrations, the models used in the self-calibration process were created by cleaning the emission down to a conservative 6$\sigma$ noise level. For the phase-amplitude round, the model was created by cleaning down to a 1$\sigma$ noise level to include as much flux as possible in the model for the amplitude calibration. All rounds of self-calibration (phase-only and phase-amplitude) were performed for the short-baseline observations of all spectral settings and the combined data of the Band 7 observations. For the combined datasets, no self-calibration was performed for the Band 3 spectral settings, as many solutions were flagged. Following the same reasoning, self-calibration of the combined Band 4 observations was stopped after three rounds of phase-only self-calibration. Table \ref{tab:SelfCal-S/N} lists the starting and final $S/N$-ratio of the continuum emission in all spectral settings. \\
\indent We used the CASA task \textsc{tclean} to extract spectra and search for molecular emission of known transitions in the Cologne Database for Molecular Spectroscopy (CDMS; \citealt{MullerEA01,MullerEA05}). Before imaging, we performed a continuum subtraction using the \textsc{uvcontsub}-task of \textsc{CASA}, selecting line-free regions and using a fit-order of unity. The images were created using the `Briggs' weighting scheme and robust parameters of +0.5 in the case of strong detections, and +2.0 in the case of weak detections. As the velocity resolution changes between the different spectral settings and programs, the molecular transitions were imaged with resolutions of 0.30~km~s$^{-1}$ in Bands 3 and 4, and 0.15~km~s$^{-1}$ in Band~7. For transitions located in continuum spectral windows, which were observed in the `Frequency Domain Mode', we used velocity resolutions of 2.0~km~s$^{-1}$, 1.20~km~s$^{-1}$, and 1.00~km~s$^{-1}$. \\

\section{Molecular emission}\label{sec:Molecules}
In case of a detected molecular species, we used a Keplerian mask (see \citealt{KeplMask}) to clean the emission down to a level of 3$\times$ the noise in the dirty image. The Keplerian mask, using an outer radius of 2.5" for the molecular emission, was defined using a central mass of 2.39~M$_\odot$ (host star plus companion mass; \citealt{VioqueEA25}), a distance of 159.26~pc, and a disk inclination and position angle of, respectively, 28.3\degree and 162.5\degree (see Section \ref{sec:IncPA} for more details). Furthermore, we used the Python package \textsc{GoFish} \citep{GoFish} to confirm the detection of weak molecular emission. By accounting for the Keplerian rotation of the disk, \textsc{GoFish} can improve the $S/N$-ratio for lines by aligning and stacking spectra taken from different sides of the disk (see also \citealt{YenEA16}). Weakly detected species and non-detections are discussed in Section \ref{sec:WNDetections}. \\
\indent Figure \ref{fig:KeyGallery} displays the key detected molecular species for our analysis, while \ref{fig:Gallery} contains the expanded molecular inventory of the disk around HD 142527. Using the new ALMA observations, we report strong detections of 6 molecular species: two transitions of \ce{C^{17}O}, two additional transitions of \ce{CS}, one transition of SO, three transitions of \ce{CN}, four transitions of \ce{C_2H}, and one additional transition of \ce{H_2CO}. Molecular transitions of \ce{C^{17}O}, \ce{SO}, \ce{CN}, and \ce{C_2H} are detected for the first time in the disk of HD~142527. Furthermore, we report the detection of one additional transition of \ce{HCO^+} ($J$=1-0) that was found in archival Band~3 observations. The integrated intensity (or moment-0) maps of these transitions are shown, together with previously observed molecular species \citep{CasassusEA13,vdPlasEA14,TemminkEA23}, in Fig. \ref{fig:Gallery}. Further information on the observed transitions, their line properties (all taken from CDMS), and peak fluxes are listed in Table \ref{tab:Detections}. \\
\indent As can be seen in both Fig. \ref{fig:KeyGallery} and Fig. \ref{fig:Gallery}, the molecular emission, except for the more abundant \ce{CO} isotopologues, is dominated by asymmetric features. However, these molecular asymmetries are distributed differently throughout the disk. While the emission of \ce{C^{17}O} and the \ce{HCO^+} $J$=1-0 transition is co-spatial with the continuum emission, the emission of the \ce{CS} transitions (at both similar and different frequencies) is most prominent in the south-western side of the disk. Even though the emission of \ce{H_2CO}, \ce{CN}, and \ce{C_2H} peaks in a similar region as the \ce{CS} transitions, their morphology displays a spiral-like feature. We discuss these different asymmetric features and their potential origins in Section \ref{sec:MolAsym}. \\
\indent While the molecular emission is dominated by asymmetries, we highlight that many molecular transitions display Keplerian signatures in the form of emission rings. Although weak, these rings are visible in the molecular emission of the \ce{CS} transitions (except for the $J$=10-9 transition), the stacked images of \ce{CN} and \ce{C_2H} transitions, and that of the \ce{H_2CO} $J$=5$_{1,5}$-4$_{1,4}$ transition. This suggests that, even though the asymmetries dominate the emission, the molecules are fully distributed along the disk's azimuth and are (partially) captured on Keplerian orbits. \\
\indent As the individual transitions of \ce{CN} and \ce{C_2H} are close together in upper level energies and Einstein-A coefficients (see Table \ref{tab:Detections}), we expect that they trace the same gas. Therefore, we do not expect that the molecular emission in the stacked images is subject to any blurring effects caused by, for example, the stacked transitions originating from different emitting layers.

\section{Analysis and discussion} \label{sec:Disc}
\subsection{Inclination and position angle} \label{sec:IncPA}
\begin{figure*}[ht!]
    \centering
    \includegraphics[width=\textwidth]{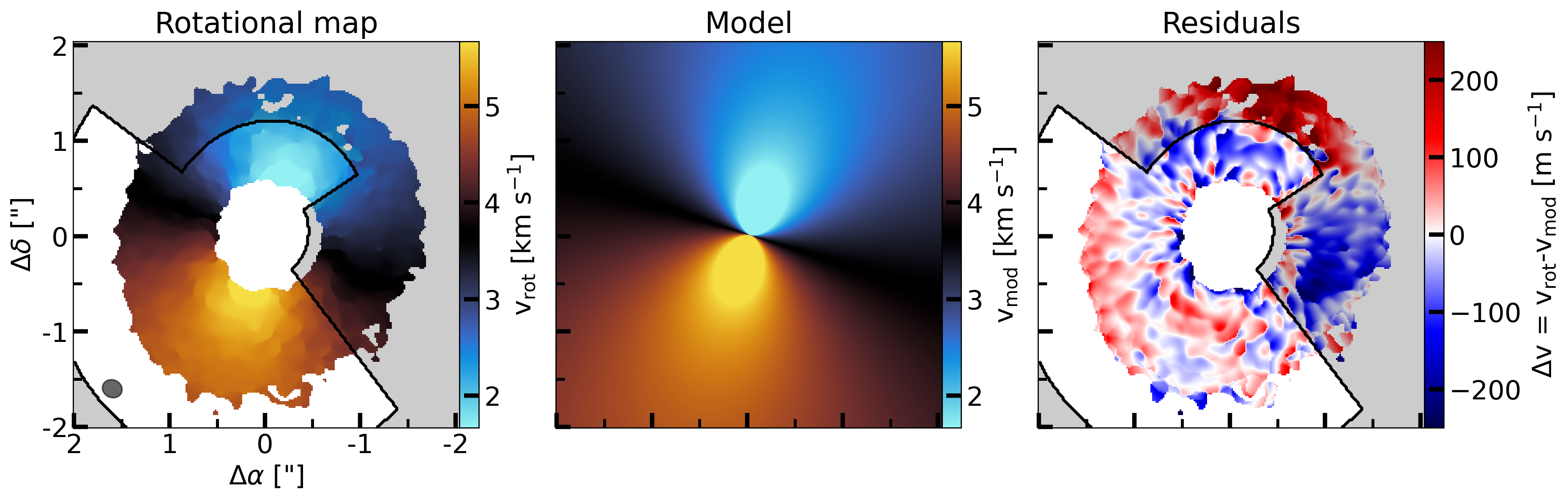}
    \caption{Rotational velocity map, resulting model, and residuals of the \ce{C^{17}O} $J$=3-2 transition, imaged with a velocity resolution of 0.15~km~s$^{-1}$. The used mask is shown in black. The greyed-out regions have not been used in the fitting process.}
    \label{fig:C17O-Eddy}
\end{figure*}
Using the \ce{C^{17}O} $J$=3-2 transition, we reanalyse the inclination, position angle, and system velocity ($v_\mathrm{lsr}$) for the disk of HD~142527. As the \ce{C^{17}O} is a rare isotopologue, this transition has the benefit of the emission originating from close to the disk's midplane. Therefore, we fit the rotational velocity maps, created with the \textsc{bettermoments} code (clipping all $<$5$\sigma$ data; \citealt{bettermoments}), using the thin disk model implemented in \ce{eddy} \citep{eddy}. As the inclination and stellar mass are degenerate, we fix the stellar mass to the total summed mass of the host star and the companion ($M_\mathrm{tot}$=2.39~M$_\odot$; \citealt{VioqueEA25}). \\
\indent While fitting the rotational velocities, we noticed strong super-Keplerian residuals in the northern side of the disk. The western side of the disk was, on the other hand, dominated by sub-Keplerian residuals (see the right panel of Fig. \ref{fig:C17O-Eddy}). To ensure these regions did not impact the Keplerian models, we excluded them, through trial and error, when creating the rotational velocity maps. Figure \ref{fig:C17O-Eddy} shows the full rotational map, model and residuals for completion. This fit yields an inclination of $i$=28.3\degree, a position angle of $PA$=162.5\degree, and a velocity of $v_\mathrm{lsr}$=3.67~km~s$^{-1}$. Uncertainties are found to be very small, well below $<$1\%, for all parameters. The derived value for the inclination closely represents that found by \citet{FukagawaEA13} ($i\sim$27\degree and $PA\sim$160\degree), yet it is significantly lower compared to that derived by \citet{BohnEA22}, following from their lower assumed stellar mass of 1.75~M$_\odot$.

\subsection{Brightness temperature} \label{sec:BT}
\begin{figure*}[ht!]
    \centering
    \includegraphics[width=0.9\textwidth]{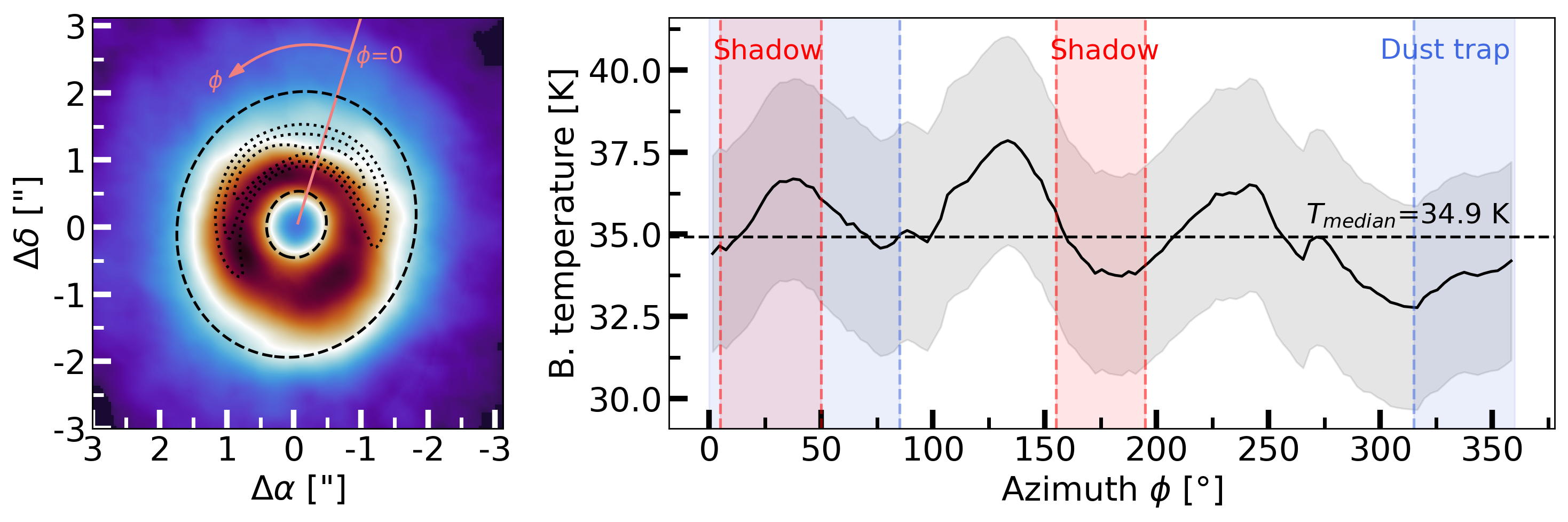}
    \caption{Brightness temperature map of the \ce{^{13}CO} $J$=2-1 transition (left panel) and the azimuthal peak temperature profile between 0.5" and 2.0" (right panel). The location of the shadows is indicated by the red shaded areas, whereas the location of the strongest emission of the dust continuum is indicated by the blue shaded areas.}
    \label{fig:13CO-BT}
\end{figure*}
One potential explanation for the different observed molecular distributions could be azimuthal variations in the temperature. We investigate the brightness temperature of the \ce{^{13}CO} $J$=2-1 transition. Since the \ce{^{12}CO} emission is dominated by the various spirals and affected by the continuum absorption \citep{ChristiaensEA14,GargEA21}, and the \ce{C^{18}O} emission was found to be only moderately optically thick ($\tau_\mathrm{\ce{C^{18}O}}\sim$0.7; \citealt{TemminkEA23}), the optically thick \ce{^{13}CO} emission yields the best opportunity to directly study the gas temperature across the disk's radial and azimuthal extent. \\
\indent The left panel of Fig. \ref{fig:13CO-BT} shows the brightness temperature map, while the azimuthal peak temperature profile, taken between radial distances of 0.5" and 2.0" with azimuthal increments of $\delta\phi$=3\degree, is shown in the right panel. Uncertainties on the integrated intensity map have been estimated following the method described in \citet{LeemkerEA22}. The red shaded areas indicate approximately the location of the shadows seen in the scattered light, while the blue shaded areas indicate the azimuthal increment in which the continuum is the strongest. We find that the temperature of the \ce{^{13}CO} emission is fairly constant, with maximum differences of $\sim$5 K, and we obtain a median peak brightness temperature of $T_\mathrm{med,^{13}CO}\sim$34.9 K. While differences of $\Delta T\sim$5~K may have localised impacts on some chemical reactions, we note that these differences occur over the azimuthal peak regions of \ce{C_2H}, \ce{CN}, and \ce{CS} ($\delta\phi\sim$175-300\degree). Therefore, azimuthal temperature variations should not be the cause of the observed molecular asymmetries.

\subsection{Molecular asymmetries and structures} \label{sec:MolAsym}
In the following sections, we discuss the various asymmetries and substructures seen in the molecular emission. We tackle different subsets of molecular species and propose scenarios that may explain the origins of these morphologies.

\subsubsection{Enhanced surface density: \ce{C^{17}O} and \ce{HCO^+}} \label{sec:OverD}
The emission of both \ce{C^{17}O} transitions is strikingly coincidental with the dust trap, which is in contrast with that of the main isotopologues of \ce{CO}, which are distributed throughout the disk's full azimuth. As opposed to the (moderately) optically thick emission from the main isotopologues, the \ce{C^{17}O} emission is very likely optically thin, assuming that the isotopic ratio of \ce{^{18}O}/\ce{^{17}O}$\sim$3.6 for the local interstellar medium \citep{Wilson99} holds locally in the disk. While optically thick emission traces the gas temperature (see Section \ref{sec:BT}), the optically thin \ce{C^{17}O} emission will trace the surface density of the gas. As the emission is coincidental with the continuum emission, this suggests that there could be an enhancement in the surface density at the location of the dust trap. \\
\indent There is one striking difference between the emission of the \ce{C^{17}O} $J=1-0$ and $J$=3-2 transitions. While the overall weaker $J$=1-0 transition is only located at the exact location of the dust trap, the stronger $J$=3-2 is more azimuthally extended and, crucially, lacks emission in the northern side of the disk. One potential explanation for this observed decrement is continuum oversubtraction effects. Such effects can arise in two different ways \citep{BoehlerEA17,WeaverEA18,NazariEA23}. First, in the case of optically thick line emission, the molecules block the thermal continuum emission originating from the disk's midplane. In this case, continuum oversubtraction techniques remove line emission instead of continuum emission. Second, in the case of optically thin line emission but optically thick continuum emission, the optically thick continuum will block emission originating from the disk's backside. This second scenario can effectively remove up to half the line flux. As the disk has a relatively low inclination ($i$=28.3\degree), the decrement in emission may be the result of this second scenario. The lack of such a decrement in the $J$=1-0 transition may be due to a difference in optical depth between the Band~3 (3~millimetre) and Band~7 (0.9~millimetre) continuum observations. \citet{GuidiEA22} showed that the continuum emission of the disk around HD~163296 is still optically thin ($\tau<$1) in the Band~3 observations, while it became optically thick ($\tau>$1) in Band~7. To gain insights into the optical depth of the outer disk, we have computed both the continuum brightness temperature and the spectral index, using the peak fluxes in each Band (details are given in Appendix \ref{sec:Dust-OD}). In the case of optically thick emission, the peak brightness temperature of $T_\mathrm{b,dust}\sim$30~K suggests that the temperature of the midplane is slightly higher than that of \ce{CO} desorption at $\sim$20-25~K. The dust trap being too warm for \ce{CO} to be frozen out is in agreement with the RADMC-3D model by \citet{TemminkEA23}, which suggests that the snowline is located at the outer edge of the trap (see Figure \ref{fig:Gallery}). A spectral index of $\alpha\sim$2.36-2.74 disagrees with optically thin emission, which requires $\alpha>$3. Instead, it is consistent with optically thick emission (requiring $\alpha\sim$1.6-2.5; \citealt{TazzariEA21}) or grain-growth. Overall, we consider optically thick continuum a valid assumption, yet detailed modeling of the dust optical depth at both 3~mm and 0.9~mm is required to confirm whether the continuum emission is optically thick in Band 7 and not in Band 3, and, therefore, can explain the morphological difference in the \ce{C^{17}O} $J$=1-0 and $J$=3-2 transitions. \\
\indent Similar to the \ce{C^{17}O} emission, the \ce{HCO^+} $J$=1-0 transition is co-spatial with the dust continuum. A simple explanation can be found in the main gas-phase formation reaction governing the \ce{HCO^+} abundances in disks \citep{HK73,LeemkerEA21}:
\begin{align}
    \mathrm{\ce{CO} + \ce{H_3^+}} & \rightarrow \mathrm{\ce{HCO^+} + \ce{H_2}}. \label{eq:HCO+-Form}
\end{align}
This reaction involves gas-phase \ce{CO} and, therefore, the same enhanced surface density as seen for the \ce{C^{17}O} transitions may be the cause of the \ce{HCO^+} $J$=1-0 transition being co-spatial with the dust continuum. \\
\indent One clear difference between the \ce{HCO^+} $J$=1-0 and $J$=4-3 transitions is that the former is dominated by emission that peaks co-spatially at the location of the dust trap, while the latter reveals emission along the disk's full azimuth. However, the $J$=4-3 transition is dominated by a bright central spot, which was also seen for the $J$=8-7 transition \citep{TemminkEA23}. \citet{CasassusEA13} proposed that this bright spot is connected to the outer disk and must be the result of inflowing material. It is, however, also possible that the emission connecting the bright inner spot with the outer disk is due to beam smearing effects. The bright emission spot being seen in the $J$=4-3 and $J$=8-7 transition, but not in the $J$=1-0 transition, may be due to excitation effects, as the warmer temperatures in the inner regions excite the higher energy transitions more easily. Furthermore, \ce{HCO^+} has a slower photodissociation rate compared to other molecules \citep{HeaysEA17} and, therefore, once formed in the cavity, it can survive for an extended period of time. An alternative explanation could be X-ray flares, as such flares can affect gas-phase cations (including \ce{HCO^+}) and temporarily enhance their abundances \citep{WaggonerEA22,WaggonerEA23}. The archival \ce{HCO^+} observations consist of a single execution block, and investigation of the individual scans yields no indications of a potential X-ray flare. Additional observations of this transition, covering a larger time span, are needed to conclude whether X-ray flares may indeed be the cause of this bright emission spot.

\subsubsection{Spirals and late infall: \ce{H_2CO}, \ce{C_2H}, and \ce{CN}} \label{sec:Mol-Infall}
\begin{figure*}[ht!]
    \centering
    \includegraphics[width=\textwidth]{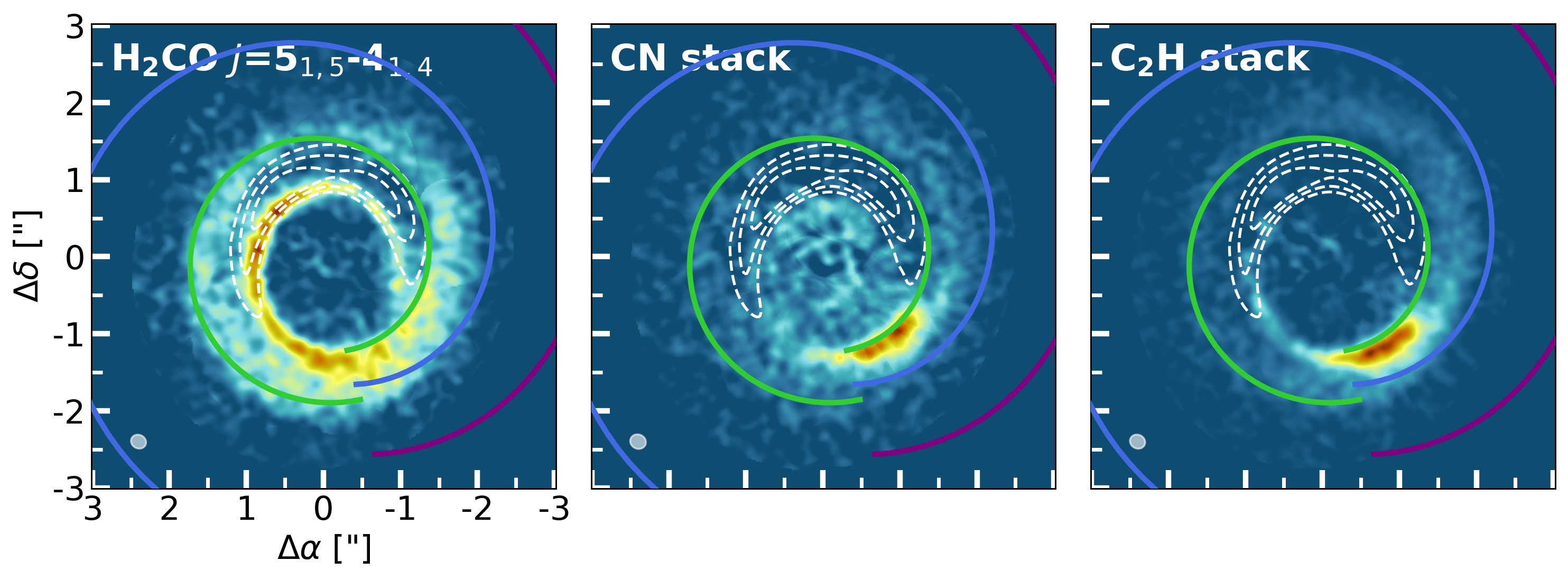}
    \caption{Integrated intensity maps of the \ce{H_2CO} $J$=5$_{1,5}$-4$_{1,4}$ transition and the stacked \ce{CN} and \ce{C_2H} transitions. Overlaid are the traced spiral features from the \ce{^{12}CO} brightness temperature channel maps (see Appendix \ref{sec:Spirals}). The white, dashed contours indicate the continuum emission at flux levels of 25\%, 50\%, and 75\%.}
    \label{fig:SpiralMolecules}
\end{figure*}
We now turn to the distribution of \ce{H_2CO}, \ce{C_2H}, and \ce{CN}, all of which peak opposite to the dust trap. Since temperature variations are excluded, another process must be at work. A possible explanation lies in the presence of spiral arms, of which large-scale structures have already been identified in the emission of \ce{^{12}CO} and \ce{^{13}CO} \citep{ChristiaensEA14,GargEA21,WoelferEA23}. However, similar structures were not seen in the emission of other molecular species, with, for example, \citet{vdPlasEA14} reporting that no counterpart of these spiral structures was seen in the \ce{CS} $J$=7-6 and \ce{HCN} $J$=4-3 transitions. Our new ALMA observations create a new perspective, as spiral-like features are clearly visible in the emission of the \ce{H_2CO} $J$=5$_{1,5}$-4$_{1,4}$ transition, the stacked transitions of \ce{C_2H} and \ce{CN} (see Fig. \ref{fig:SpiralMolecules}), and even the selected channel maps of the stacked \ce{c-C_3H_2} transitions (see bottom row of Fig. \ref{fig:CM-WT}). Figure \ref{fig:CM-H2CO} displays the channel maps of the \ce{H_2CO} $J$=5$_{1,5}$-4$_{1,4}$ transition together with overlaying contours of the dust continuum. These channels reveal that the spiral-like features are real and not an artefact due to, for example, the optically thick continuum. Additionally, we have distinguished between the contribution from the disk and that of the spiral-like feature. Using these new observations as guidance, we report that weak hints for similar structures can also be seen in the \ce{CS} $J$=7-6 and the \ce{HCN} and \ce{HCO^+} $J$=4-3 transitions (see Fig. \ref{fig:Gallery}).  \\
\indent The first question to address is whether the observed spiral-like features in the emission of the \ce{H_2CO}, \ce{CN}, and \ce{C_2H} transitions align with the known spiral arms observed in the \ce{^{12}CO} $J$=2-1 transition. To test the alignment, we have traced the spiral arms in the \ce{^{12}CO} (see Section \ref{sec:Spirals} for more details) and overlaid the resulting spirals on top of the integrated intensity maps of \ce{H_2CO}, \ce{CN}, and \ce{C_2H} (see Fig. \ref{fig:SpiralMolecules}). The traced spirals appear to align rather well with those seen in the other molecular species. Offsets between the observations may be due to the lower spatial resolution of the \ce{^{12}CO} emission. \\ 
\indent As the molecular emission now appears to be directly intertwined with the presence of these spiral features, another question must be posed: how are these structures influencing the observable chemistry in the disk of HD~142527? Both \ce{CN} and \ce{C_2H} are linked to UV-chemistry, following the need for atomic carbon in their gas-phase formation pathways. Similarly, \ce{CN} is also known to be the photodissociation product of \ce{HCN} \citep{SD95}. A direct role of UV-driven chemistry in setting the observable emission morphologies is, however, not evident. As the molecular emission is found at a large distance ($>$150~au) with respect to the host star, and small dust grains are present along the disk's entire azimuth (see Fig. \ref{fig:IntroOverview}), we consider it unlikely that UV radiation dominates the observable chemistry at one specific location outside of the shadows. \\
\indent Aside from UV-chemistry, a locally varying carbon-to-oxygen ratio (C/O-ratio) can also affect the formation efficiency of \ce{C_2H}, \ce{CN}, and \ce{c-C_3H_2}. Models have shown that a strong reservoir of hydrocarbons, such as \ce{C_2H}, requires a C/O-ratio larger than unity \citep{BerginEA16,KamaEA16,MiotelloEA19}. The emission of \ce{CN} is similarly impacted by an increasing C/O-ratio. However, the effects are weaker compared to \ce{C_2H}: when increasing the C/O-ratio from 0.3 to 1.5, the \ce{C_2H} abundance was shown to increase up to an order of magnitude, while that of \ce{CN} only increases by a factor of two \citep{CazzolettiEA18,MiotelloEA19}. As our observed peak fluxes of \ce{C_2H} are stronger than those of \ce{CN} (see Table \ref{tab:Detections}), a locally enhanced C/O-ratio may be the cause of the observed emission. \\
\indent We propose that the observed spiral-like features are the result of low-density, late-infalling material or streamers that replenish the disk with material. In particular, in low-density environments (such as translucent clouds or the disk's atmospheric layers), not all atomic carbon will be locked up in \ce{CO} (see, for example, \citealt{vDB88}) and is, therefore, available for the formation of hydrocarbons. \citet{LesurEA15} have shown that infalling material can yield long-lived spiral arms in planet-forming disks and can result in efficient radial transport of angular momentum. Another scenario could be a locally depleted oxygen abundance following the freeze-out of oxygen-bearing species. This latter scenario is less likely, as the majority of the millimetre-sized dust grains are trapped on the opposite side of the disk. It must be noted that the southern side of the disk is not completely void of large dust grains and that the small dust grains are distributed along the disk's full azimuth. Therefore, freeze-out cannot be fully ruled out, but it is alone unlikely to explain the observed \ce{C_2H} and \ce{CN} emission peaking at a specific azimuthal location. \\
\indent The detection of \ce{H_2O}-ice confirms that freeze-out takes place in the disk of HD~142527 \citep{HondaEA09,MinEA16,TazakiEA21}. Furthermore, \citet{MinEA16} used observations of the \textit{Herschel} Photodetector Array Camera and Spectrometer (PACS) instrument to conclude that 80\% of the available oxygen atoms must be locked up in the \ce{H_2O}-ice. Since the RADMC-3D model of \citet{TemminkEA23} yields dust temperatures of 40-50~K at the location of the dust trap, it is not surprising that oxygen-bearing species such as \ce{H_2CO} (binding energy of $E_\mathrm{bind}>$3300~K; \citealt{PenteadoEA17,MinissaleEA22}) and \ce{H_2O} and \ce{CH_3OH} (both have binding energies of $E_\mathrm{bind}\gtrsim$5600~K; \citealt{FraserEA01,PenteadoEA17,MinissaleEA22}) will be frozen out. The presence of a spiral-like feature in the \ce{H_2CO} emission is, in that case, surprising, unless the infalling material or streamer is able to release ices through a weak shock. While a shock may explain the presence of \ce{H_2CO}, it does not explain the emission morphology of another known shock-tracer, \ce{SO} \citep{vGelderEA21}, which has a lower binding energy of $E_\mathrm{bind}\gtrsim$2800 K \citep{PenteadoEA17} and, therefore, should also be released from the grains. The \ce{SO} emission is further discussed in Section \ref{sec:Mol-SO}. \\
\indent We note that \ce{H_2CO}, unlike \ce{CH_3OH}, can also form efficiently in the gas. The reactions involve atomic oxygen, OH, and the radical hydrocarbons \ce{CH_3} and \ce{CH_2} \citep{FP02,AtkinsonEA06}:
\begin{align}
    \mathrm{\ce{CH_3}} + \mathrm{\ce{O}} & \rightarrow \mathrm{\ce{H_2CO}} + \mathrm{\ce{H}} \label{eq:H2CO-Form1} \\
    \mathrm{\ce{CH_2}} + \mathrm{\ce{OH}} & \rightarrow \mathrm{\ce{H_2CO}} + \mathrm{\ce{H}} \label{eq:H2CO-Form2}
\end{align}
Since both formation routes involve a hydrocarbon, a locally increased C/O-ratio will enhance the \ce{CH_3} and \ce{CH_2} abundances and may, therefore, boost the gas-phase formation of \ce{H_2CO}. Since the spiral-like features are strongest in the emission of \ce{H_2CO}, \ce{CN}, and \ce{C_2H}, while weak signatures can be found in the emission of the other carbon-bearing species \ce{HCN} and \ce{CS}, we consider the local enhancement of the C/O-ratio, following from infalling material replenishing the disk with fresh atomic carbon and potentially other carbon-rich material, as a viable scenario for the observed emission morphology of these molecular species. \\
\indent To summarise this section, we see spiral-like features in the emission of \ce{H_2CO}, \ce{CN}, \ce{C_2H}, and \ce{c-C_3H_2}, and we propose that similar features are weakly visible in those of the previously detected \ce{CS} $J$=7-6 and \ce{HCN} and \ce{HCO^+} $J$=4-3 transitions. We consider it most likely that these features are due to an influx of fresh atomic carbon, syphoned onto the disk by late infalling material or streamers along the spiral arms seen in the \ce{^{12}CO} emission. This influx of C-rich material locally increases the C/O-ratio and facilitates the gas-phase formation of hydrocarbons, \ce{CN}, and even that of \ce{H_2CO}. We consider an increasing C/O-ratio following a depleted oxygen abundance to be less likely, as this would suggest that the \ce{H_2CO} emission may trace sublimation due to the infalling material creating a weak shock at its impact location on the disk. \\
\indent To further support the scenario of infalling, atomic carbon-rich material, we highlight that (atomic) carbon-rich streamers may not be uncommon in young stellar objects (YSOs). For example, carbon-bearing molecules, such as \ce{C_2H}, \ce{c-C_3H_2}, and \ce{HC_3N}, have been detected in the streamers of the Class 0 YSO Per-emb-2 \citep{TaniguchiEA24} and the Class 1 YSO L1489 IRS \citep{TaniousEA24,TaniousEA25}. 

\subsubsection{Kinematical support of late-infall scenario} \label{sec:LI-Kin}
Low-density, infalling material may provide a potential explanation for the observed emission morphologies of \ce{CN}, \ce{C_2H}, and \ce{H_2CO}, so one may ask whether this is supported by the kinematics of the gas. Therefore, we place the kinematical analysis of \citet{GargEA21} in context with the infall model of \citet{CalcinoEA25}. \\
\indent While the foreground absorption complicates the analysis, \citet{GargEA21} note that the `S4' spiral (the green spiral in Figures \ref{fig:SpiralMolecules} and \ref{fig:12CO-Spirals}) appears to be super-Keplerian in the red-shifted regions (south-eastern side of the disk, see Figure \ref{fig:C17O-Eddy}), while sub-Keplerian in the blue-shifted regions (north-western regions). They provide a possible explanation for this variation in pattern speed: vertically ascension of the spiral arms. Intuitively, the assumption can be made that the ascension occurs outwards in the radial direction and could, therefore, support our proposed scenario of infalling material setting the chemistry in the disk of HD~142527. As the spiral is most prevalent in the \ce{H_2CO} emission, higher spectral resolution observations (ideally $\sim$0.1 km~s$^{-1}$) of \ce{H_2CO} transitions are needed to investigate the kinematics of the molecular spiral. \\
\indent The models by \citet{CalcinoEA25} reveal that infalling material can simultaneously yield both stationary and super-Keplerian spiral features. As the `S4' spiral is (partially) rotating super-Keplerian, this may verify the notion by \citet{GargEA21} that the spiral is vertically ascending or infalling. Finally, we highlight that the models suggest that these spirals can persist in the disks for over $>$10$^4$ years.

\subsubsection{\ce{H_2CO} formation timescale delay}
As shown in Fig. \ref{fig:SpiralMolecules}, the \ce{H_2CO} $J$=5$_{1,5}$-4$_{1,4}$ is azimuthally offset from the peak location of the \ce{CN} and \ce{C_2H} emission. We propose that this may be due to a delay in the gas-phase formation of \ce{H_2CO} through Equations \ref{eq:H2CO-Form1} and \ref{eq:H2CO-Form2}. To investigate the potential delay timescale, Fig. \ref{fig:Deprojected} displays the deprojected integrated intensity maps of the stacked \ce{C_2H} and \ce{H_2CO} $J$=5$_{1,5}$-4$_{1,4}$ emission. In both images, we have visually indicated the peak location of the \ce{C_2H} emission, which has a radial extent between 215 and 275 au and an azimuthal extent of 210$\degree$ to 255$\degree$ or 1/8$^\mathrm{th}$ of the disk's azimuth. We further note that \ce{H_2CO} emission is stronger outside this defined box. At these radial locations, the Keplerian orbital timescales of the disk are $\tau_\mathrm{215au}$=2041 and $\tau_\mathrm{275au}$=2952 years. By taking the average of 1/8$^\mathrm{th}$ of these timescales - thus, accounting for the azimuthal extent of the \ce{C_2H} - the peak \ce{H_2CO} emission is delayed by $\tau_\mathrm{\ce{H_2CO},delay}$=312 years. This timescale holds under the assumption of Keplerian emission, while spiral-like features in reality move on slightly sub- or super-Keplerian speeds. As previous studies have shown that the spiral features leave velocity residuals of $<$1 km~s$^{-1}$ (see, for example, \citealt{TeagueEA19}), such velocity differences do not change the estimated timescale significantly, and, therefore, a Keplerian orbit can be used for a ballpark estimate. \\
\indent To investigate whether this difference is due to a delay in the gas-phase formation of \ce{H_2CO}, we have calculated the formation timescale of Reactions \ref{eq:H2CO-Form1} and \ref{eq:H2CO-Form2} for different gas densities ($n(\mathrm{\ce{H_2}})$) and fractional abundances ($X(\mathrm{\ce{CH_x}}/\mathrm{\ce{H_2}}$) of \ce{CH_3} and \ce{CH_2}, the expected limiting reactants. Using the Kinetic Database for Astrochemistry (KIDA; \citealt{WakelamEA12}), we retrieve an average reaction rate coefficient of 1.1$\times$10$^{-10}$ cm$^3$ s$^{-1}$ (at a temperature of $T$=298 K) for Equation \ref{eq:H2CO-Form1} and a reaction rate coefficient of 3.0$\times$10$^{-10}$ cm$^3$ s$^{-1}$ (at temperatures of $T$=10-280 K) for Equation \ref{eq:H2CO-Form2}. Since the temperature dependence of these radial-atom reactions is expected to be weak, we expect that these rate coefficients hold for the disk. Furthermore, we explore the reaction rates for gas densities of 10$^6$ to 10$^9$ cm$^{-3}$, as the gas density of the disk depends on the vertical emitting layer of the disk. Similarly, we use strongly enhanced values of 10$^{-6}$ to an expected value of 10$^{-10}$ for the fractional abundance. A thermochemical model of a massive disk, such as HD 142527, suggests that fractional abundance of $\sim$10$^{-8}$ and $\sim$10$^{-10}$ can be reached for, respectively, \ce{CH_2} and \ce{CH_3} in their respective emitting layers (M. Leemker, priv. comm.). A formation timescale of the same order of magnitude as the above derived $\tau_\mathrm{\ce{H_2CO},delay}$=312 years can be reached for any combination of $n(\mathrm{\ce{H_2}})$=10$^Y$ cm$^{-3}$ and $X(\mathrm{\ce{CH_x}}/\mathrm{\ce{H_2}})$=10$^{-Y}$, where $Y$ is the same number. \\ 
\indent Since lower densities are expected for emitting layers higher up in the disk, this suggests that enhanced fractional abundances are needed for these layers. As the above suggested scenario of infalling material is a local effect, it is possible that the fractional abundances, similarly to the C/O-ratio, are locally enhanced. Therefore, we deem it likely that the azimuthal differences between the \ce{H_2CO} $J$=5$_{1,5}$-4$_{1,4}$ peak emission and the stacked \ce{CN} and \ce{C_2H} can be due to a lag in the \ce{H_2CO} gas-phase formation timescales. Furthermore, the lag may also be introduced by the slow formation rate of the precursors \ce{CH_2} and \ce{CH_3}. In particular, \ce{CH_3} forms through the, in comparison with the formation routes of \ce{C_2H} and \ce{CN}, slow radiative association of \ce{CH_3^+}+\ce{H_2} that forms \ce{CH_5^+} and the subsequent dissociative recombination into neutral hydrocarbons \citep{MN85}. This analysis shows that time-dependent chemistry can play an important role in setting the observable molecular composition in planet-forming disks and that systems with (molecular) asymmetries and/or signatures of infalling material yield the unique opportunity to study these processes.

\subsubsection{Azimuthal or vertical variations: CS}
Including the new Cycle 10 and 11 observations, four molecular transitions of \ce{CS} have been detected in the disk of HD 142527  (compared to \citealt{vdPlasEA14} and \citealt{TemminkEA23}): $J$=2-1 ($E_\mathrm{up}\sim$7 K), $J$=3-2 ($E_\mathrm{up}\sim$14 K), $J$=7-6 ($E_\mathrm{up}\sim$66 K), and $J$=10-9 ($E_\mathrm{up\sim}$129 K). As shown in the second row of Fig. \ref{fig:Gallery}, the emission morphology changes between the transitions. While all four transitions exhibit the strongest emission in the western side of the disk, the $J$=2-1 and $J$=3-2 are much more ring-like. The $J$=7-6 transition shows weak hints of the spiral-like feature that is also seen in the emission of \ce{H_2CO}, \ce{C_2H}, and \ce{CN}, and the $J$=10-9 transition is potentially co-spatial with the bright emission spot seen for \ce{C_2H} and \ce{CN}. \\
\indent Models by \citet{FF20} suggest that the main \ce{CS} emitting layer is located below that of \ce{^{13}CO}. As the brightness temperature of \ce{^{13}CO} was found to be fairly constant along the disk’s azimuth, with maximum variations of $\sim$5 K (see Fig. \ref{fig:13CO-BT}), we do not expect the differences between the \ce{CS} transitions to be due to azimuthal temperature variations. \\
\indent As the $J$=7-6 transition appears to exhibit weak signatures of the spiral-like feature, and the $J$=10-9 transition may be co-spatial with the bright emission spot seen in the \ce{C_2H} and \ce{CN} transitions, we propose that we are observing two vertically distinct CS reservoirs. The first reservoir comes from deep inside the disk, closer to the midplane and below the \ce{^{13}CO} emitting layer, where the emission follows a Keplerian orbit. The emission of the $J$=2-1 and $J$=3-2 transitions, and part of the $J$=7-6 transition, originates from this reservoir. The second, hotter reservoir resides higher up in the disk and is facilitated by the infalling material or streamer. Gas-phase formation reactions, such as the neutral-neutral reactions \ce{CH}+\ce{S}$\rightarrow$\ce{CS}+\ce{H} or \ce{HS}+\ce{C}$\rightarrow$\ce{CS}+\ce{H} \citep{VidalEA17}, may play a role in forming this second reservoir. Additional formation routes include the gas-phase reactions of \ce{S^+} with hydrocarbons of the form \ce{CH_x} and \ce{C_yH} (where $x$=1-4 and $y$=2-3), which produce \ce{HCS^+}, \ce{CS^+}, \ce{HC_3S^+}, and \ce{C_2S^+}, which may form neutral S-bearing species, such as \ce{CS} and \ce{H_2CS}, after recombination reactions with electrons \citep{LeGalEA19}. These reactions all require the presence of \ce{S^+}, \ce{S}, \ce{HS}, and atomic carbon or hydrocarbons. While the abundances of the former three have not been established for the disk of HD 142527, the presence of atomic carbon and hydrocarbons can be due to infalling material or a streamer that replenishes the disk with fresh atomic carbon-rich material, following the same reasoning as in Section \ref{sec:Mol-Infall}. \\
\indent This scenario may also explain the observed \ce{H_2CS} $J$=10$_{1,10}$-9$_{1,9}$ transition. As can be seen in the middle panel of Fig. \ref{fig:CM-WT}, the \ce{H_2CS} transition peaks in a similar region to those of \ce{C_2H} and \ce{CN}, and we expect that the observed emission is facilitated by the gas-phase reactions discussed in the previous paragraph. Furthermore, as both the \ce{CS} $J$=10-9 and the \ce{H_2CS} $J$=10$_{1,10}$-9$_{1,9}$ have high upper level energies of, respectively, $E_\mathrm{up}$=129.3 K and $E_\mathrm{up}$=102.4 K, it is likely that the emission comes from a high atmospheric layer of the disk, especially as disks have steep temperature gradients (see, for example, \citealt{PanequeEA23}). \\
\indent Finally, we note that the \ce{C^{34}S} $J$=7-6 transition also appears to be co-spatial with the \ce{C_2H} and \ce{CN} transitions (see top panel of Fig. \ref{fig:CM-WT}). Being a rarer isotopologue of \ce{CS}, this may further agree with a second, strong reservoir of \ce{CS} formed by infalling atomic carbon-rich material. 

\subsubsection{A potential shock: SO} \label{sec:Mol-SO}
\begin{figure*}[ht!]
    \centering
    \includegraphics[width=\textwidth]{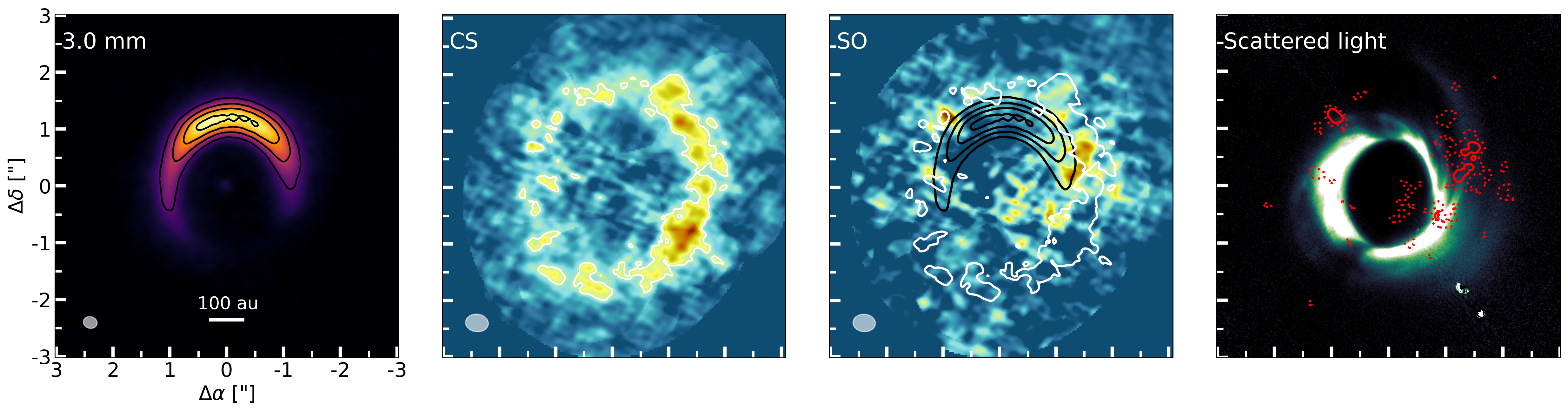}
    \caption{Continuum emission (left panel), integrated intensity maps of the CS $J$=2-1 and \ce{SO} $J$=3$_2$-2$_1$ transitions (middle panels), and the scattered light image (rightmost panel). The black contours indicate the continuum emission at 25\%, 50\%, 75\%, and 95\% the peak flux, whereas the white contours indicate the 3$\sigma$ \ce{CM} emission. In the scattered light image, the red contours indicate the \ce{SO} emission at 2$\sigma$ (dotted) and 3$\sigma$ (solid) levels.}
    \label{fig:SO-Comparison}
\end{figure*}
While a clear scenario can be put forward to explain the observed emission of \ce{H_2CO}, \ce{CS}, \ce{C_2H}, and \ce{CN}, this is not the case for \ce{SO}. We report the detection of a single, intrinsically weak ($\log_{10}(A_\mathrm{ij})\sim$-4.94) \ce{SO} transition, $J$=3$_2$-2$_1$, that was observed in both spectral settings of the Band 3 observations. One was observed with high spectral resolution, while the other was placed in a continuum spectral window. Using the high spectral resolution observation, we note that the \ce{SO} emission originates from a radial distance between the continuum emission and the \ce{CS} emission (see Fig. \ref{fig:SO-Comparison}). Furthermore, the \ce{SO} emission is not azimuthally co-spatial with other observed molecular species. \\
\indent As previously mentioned, \ce{SO} is a known shock tracer (see, for example, \citealt{vGelderEA21}). The observed emission of \ce{SO} in the disks of AB Aur and MWC 480 has previously been linked to observed gaseous spiral arms \citep{SpeedieEA25,ZagariaEA25}. In the case of HD~142527 this may be most evident once the \ce{SO} observations are compared with those in the scattered light. As shown in the right panel of Fig. \ref{fig:SO-Comparison}, there may be a connection between the observed spirals and the \ce{SO} emission. However, as the molecular emission is weak with respect to other molecular species, this can not be directly confirmed. The emission may, therefore, also be just originating from the disk. Observations of stronger \ce{SO} transitions, such as the $J$=7$_8$-6$_7$ ($E_\mathrm{up}$=81.2, $\log_{10}\left(A_\mathrm{ul}\right)$=-3.3023) and $J$=8$_8$-7$_7$ ($E_\mathrm{up}$=87.5, $\log_{10}\left(A_\mathrm{ul}\right)$=-3.2852) transitions in Band 7, are needed to investigate whether the molecular emission is caused by a shock, if it is originating from the disk itself, or if another scenario is at play. Additionally, as \ce{SO} - in combination with \ce{CS} - can be used as a proxy for the \ce{C}/\ce{O}-ratio, observations of the stronger transitions may yield additional information on this crucial elemental ratio. However, the observation of this inherently weak observed \ce{SO} transition is, on its own, not sufficient to make conclusive statements on the \ce{C}/\ce{O}-ratio from the observed \ce{CS} and \ce{SO} abundances.

\subsection{Origins of the spirals and dust trap}
Spirals are thought to form through disk interactions with heavy, embedded companions (see, for example, \citealt{Kley99,DongEA15,ZhuEA15,BZ18}), gravitational instability (see, for example, \citealt{CossinsEA09,SpeedieEA24}), and even shadows (see, for example, \citealt{MontesinosEA16,SB24,ZhangEA25}). As the system of HD 142527 has a known stellar companion and shadows due to a misaligned inner disk, none of these mechanisms can immediately be ruled out.  \\
\indent An alternative explanation for the continuum asymmetry and spiral features (see also \citealt{LesurEA15}) observed in HD~142527 can be found in the works of \citet{BaeEA15} and \citet{KuznetsovaEA22}. Their models suggest that protostellar or anisotropic infall may generate a Rossby Wave Instability (RWI; \citealt{LovelaceEA99,LiEA00}). The RWI, in turn, can give rise to vortices, which may trap the large dust grains \citep{LyraEA09,CrnkovicEA15,OK17,RaettigEA21,RegalyEA21} and lead to the large asymmetric structures seen in the continuum emission, as may be the case for the disks of Oph-IRS 48 and HD 142527. This may, therefore, have given rise to enhanced surface density proposed for the emission of, for example, the \ce{C^{17}O} transitions, which is consistent with the dust trap. As the spirals observed in the molecular emission, most prominently in that of \ce{H_2CO}, trace the inner edge of the dust trap (see Fig. \ref{fig:SpiralMolecules} and Fig. \ref{fig:CM-H2CO}), we propose that the dust trap of HD 142527 may be the result of a vortex caused by the late infalling material. \citet{StadlerEA26} also recently proposed that infalling material may be the cause of the asymmetric dust feature seen in the disk of HD~34700A. 

\subsection{Future work and studies}
The disk of HD 142527 yields unique insights into the chemistry of disks and a potential connection to infalling material, but not all molecular species tell a clear story. As mentioned in Section \ref{sec:Mol-SO}, observations of stronger transitions of \ce{SO} are needed to investigate the shocks induced by the spiral arms. Furthermore, given the low resolution, sensitivity, and image fidelity of the old Cycle 0 observations due to the smaller number of ALMA antennas, re-observing the \ce{CS} $J$=7-6, \ce{HCN}, and \ce{HCO^+} $J$=4-3 transitions will be crucial for investigating the role of this infalling material in setting the morphology of these species. Additionally, re-observing the \ce{HCO^+} $J$=4-3 transitions may reveal insights into the origins of the observed central bright spot and whether it was caused by an X-ray flare. Similarly, higher spatial resolution observations of the \ce{CS} $J$=10-9 transition, or even targeting any other \ce{CS} transition with high upper level energies, will reveal whether this transition is fully co-spatial with the observed \ce{C_2H} and \ce{CN} emission and will, therefore, confirm whether the origins of these species are all related to the potentially infalling material. \\
\indent HD 142527 is not the only disk with spiral arms seen in \ce{CO} gas emission (see, for example, \citealt{RosottiEA20,CasassusEA21,TeagueEA21,WoelferEA21,WoelferEA23}. To investigate how important and widespread our proposed scenario is for planet-forming disks, a larger sample of disks with spiral arms needs to be studied with high-resolution and high-sensitivity ALMA observations. These studies should target species that have been key in our analysis, such as \ce{H_2CO}, \ce{C_2H} (as an important C/O-ratio tracer), \ce{CN}, and \ce{CS}, but also \ce{SO} and \ce{CH_3OH}, to investigate (and potentially rule out) the role of shocks. Thus far, \ce{C_2H} has been observed in only a few of the Herbig disks studied by \citet{StapperEA22} and \citet{StapperEA24} (see, for example, \citealt{BoothEA26}), leaving a large sample open for follow-up studies. 

\section{Summary} \label{sec:Summary}
In this work, we have investigated the peculiar carbon-rich chemistry of the disk around the young star HD 142527 using new ALMA observations. Our main conclusions can be summarised as follows:
\begin{itemize}
    \item We have greatly expanded the known molecular inventory of the HD 142527 disk by detecting, for the first time, transitions of the key species \ce{C_2H}, \ce{CN}, and \ce{c-C_3H_2}. Furthermore, we detected additional transitions of \ce{CS}, \ce{H_2CO}, multiple isotopologues of \ce{HCO+} and more. The non-detection of \ce{CH_3OH} suggests that the dust trap is too cold for sublimation of the icy mantles to occur.
    \item We found that the brightness temperature of \ce{^{13}CO} is fairly constant along the disk's azimuth, suggesting that there are no thermal variations along the disk's surface. 
    \item The emission of the optically thin \ce{C^{17}O} $J$=1-0 and $J$=3-2, and \ce{HCO^+} $J$=1-0 transitions is co-spatial with the asymmetric dust trap, suggesting that the gas surface density has an enhancement at the location of the dust trap. The decrement observed in the northern region of the \ce{C^{17}O} $J$=3-2 transition may be due to the dust being more optically thick at higher frequencies, blocking emission from the disk's backside.
    \item We propose that the spiral-like features seen in the molecular emission of \ce{H_2CO}, \ce{CN}, and \ce{C_2H} transitions follow from infalling atomic carbon-rich material. This material locally increases the C/O-ratio and, therefore, facilitates the gas-phase formation of these molecular species. 
    \item An azimuthal offset between the peak emission of the \ce{H_2CO} and that of \ce{CN} and \ce{C_2H} can be explained by a delay of a few hundred years in the gas-phase formation timescales of \ce{H_2CO}. According to simple calculations, such a time-scale would require enhanced fractional abundances of the hydrocarbons \ce{CH_3} and \ce{CH_2} in the lower-density surface layers. Disks with (molecular) asymmetries yield, therefore, unique opportunities to study the role of time-dependent chemistry in setting the observable molecular composition.
    \item Four transitions of \ce{CS} have now been detected in the disk of HD 142527, and we propose that they trace two different reservoirs: a cold reservoir that follows a Keplerian orbit and a hotter reservoir, residing higher up in the disk, that is the result of the infalling atomic carbon-rich material. This is supported by the \ce{CS} $J$=10-9 and \ce{H_2CS} $J$=10$_{1,10}$-9$_{1,9}$ having a similar emission morphology as the \ce{CN} and \ce{C_2H}, and the high temperatures needed to excite these higher upper level energy transitions ($E_\mathrm{up}>$100 K).
    \item A single weak transition of \ce{SO} is detected that is radially and azimuthally offset from both the dust continuum and the \ce{CS} emission. We propose that this transition may potentially trace a shock caused by the spirals observed in the scattered light impacting the disk, but additional observations of other \ce{SO} transitions are needed to confirm this notion.
\end{itemize}
This work has shown that the azimuthally asymmetric disk of HD 142527 provides a unique laboratory to study the impact of the environment and, in particular, late infalling material on the observable chemistry. We advocate the need for a large study investigating the role and importance of spirals and/or potentially infalling material in setting the chemical composition of planet-forming disks. To conduct such a study, high-resolution and sensitivity observations, targeting molecular species such as \ce{H_2CO}, \ce{CS}, \ce{CN}, and \ce{C_2H}, are needed in many more disks with asymmetries and/or spiral features.

\begin{acknowledgements}
    The authors acknowledge assistance from Allegro, the European ALMA Regional Center node in the Netherlands. \\
    \indent This paper makes use of the following ALMA data: 2011.1.00318.S, 2011.0.00465.S, 2012.1.00631.S, 2013.1.00305.S, 2015.1.00805.S, 2015.1.01137.S, 2023.1.00628.S, and 2024.1.00446.S. ALMA is a partnership of ESO (representing its member states), NSF (USA) and NINS (Japan), together with NRC (Canada), NSTC and ASIAA (Taiwan), and KASI (Republic of Korea), in cooperation with the Republic of Chile. The Joint ALMA Observatory is operated by ESO, auI/NRAO and NAOJ. \\
    \indent M.T. and E.v.D. acknowledge support from the ERC grant 101019751 MOLDISK. E.v.D. also acknowledges support the Danish National Research Foundation through the Center of Excellence ``InterCat'' (DNRF150). M.B. has received funding from the European Research Council (ERC) under the European Union's Horizon 2020 research and innovation programme (PROTOPLANETS, grant agreement No. 101002188).  \\
    \indent The project leading to this publication has received support from ORP, that is funded by the European Union's Horizon 2020 research and innovation programme under grant agreement No 101004719 [ORP]. \\
    \indent This work also has made use of the following software packags that have not been mentioned in the main text: NumPy, SciPy, Astropy, Matplotlib, pandas, IPython, Jupyter \citep{Numpy,Scipy,AstropyI,AstropyII,AstropyIII,Matplotlib,pandas,IPython,Jupyter}
\end{acknowledgements}

\bibliographystyle{aa}
\bibliography{Bibliography}

%% ---

\begin{appendix}
\onecolumn 

\section{Observational details}
\begin{table}[ht!]
    \centering
    \caption{Observational details of the Cycle 10 and 11 ALMA observations.}
    \resizebox{\textwidth}{!}{%
    \begin{tabular}{c c c c c c c c c c}
        \hline\hline
        Band & Date         & No. antennas & On source time & Freq. coverage & Baselines & Mean PWV & MRS & Phase cal. & Flux/Bandpass cal. \\
             & (DD/MM/YYYY) &              & [min]          & [GHz]         & [m]       & [mm]     & ["] &            &                    \\
        \hline
        3 (\#1) & 06/01/2024 & 42 & 21.25 & 86.0-101.5 & 15.1-783.5 & 3.7 & 15.4 & J1604-4441 & J1427-4206 \\
                & 07/11/2023 & 48 & 42.37 & - & 30.9-6582.7 & 4.8 & 3.2 & J1604-4441 & J1924-2914 \\
                & 09/11/2023 & 48 & 42.32 & - & 30.9-5185.6 & 5.1 & 3.1 & J1604-4441 & J1427-4206 \\
        3 (\#2) & 27/12/2023 & 41 & 36.43 & 99.2-113.5 & 15.1-783.5 & 6.5 & 11.7 & J1604-4441 & J1517-2422 \\
                & 06/01/2024 & 42 & 36.45 & - & 15.1-783.5 & 4.1 & 12.0 & J1604-4441 & J1427-4206 \\
                & 07/11/2023 & 46 & 44.85 & - & 30.9-6582.7 & 4.9 & 2.5 & J1604-4441 & J1427-4206 \\
                & 07/11/2023 & 47 & 44.82 & - & 30.9-6582.7 & 4.9 & 2.7 & J1604-4441 & J1427-4206 \\
                & 08/11/2023 & 46 & 44.88 & - & 30.9-5185.6 & 5.1 & 2.8 & J1604-4441 & J1427-4206 \\
        4       & 06/01/2024 & 28 & 16.20 & 144.0-157.3 & 15.1-783.5 & 3.4 & 9.4 & J1604-4441 & J1427-4206 \\
                & 13/01/2024 & 45 & 16.20 & - & 15.1-500.2 & 2.8 & 12.7 & J1604-4441 & J1427-4206 \\
                & 03/12/2023 & 33 & 30.72 & - & 41.5-3083.2 & 0.9 & 3.1 & J1604-4441 & J1427-4206 \\
                & 05/12/2023 & 43 & 30.73 & - & 15.1-2516.8 & 1.5 & 4.2 & J1604-4441 & J1427-4206 \\
        7       & 26/12/2024 & 43 & 48.92 & 337.0-352.5 & 15.0-499.8 & 0.7 & 4.9 & J1604-4441 & J1517-2422 \\
                & 17/04/2025 & 47 & 39.27 & - & 15.1-1397.8 & 0.6 & 2.6 & J1604-4441 & J1924-2914 \\
                & 18/04/2025 & 48 & 39.33 & - & 15.1-1397.8 & 0.7 & 2.6 & J1604-4441 & J1256-0547 \\
                & 19/04/2025 & 47 & 39.32 & - & 15.1-1604.4 & 1.0 & 2.4 & J1604-4441 & J1256-0547 \\
        \hline
    \end{tabular}%
    }
    \label{tab:ObsDetails}
\end{table}

\section{Self-calibration: $S/N$-ratios}
\begin{table}[ht!]
    \centering
    \caption{$S/N$-ratios before and after self-calibration.}
    \begin{tabular}{c | c c | c c | c c | c c }
        \hline\hline
        & \multicolumn{2}{c}{Band 3 - Setting 1} & \multicolumn{2}{c}{Band 3 - Setting 2} & \multicolumn{2}{c}{Band 4} & \multicolumn{2}{c}{Band 7} \\
        & SB & Combined  & SB & Combined  & SB & Combined & SB & Combined \\
        \hline
        Start S/N & 147.5 & 106 & 250  & 147 & 184  & 192 & 90  & 170 \\
        End S/N   & 552   & -   & 919 & -    & 1808 & 514 & 483 & 672 \\
        \hline
    \end{tabular}
    \label{tab:SelfCal-S/N}
\end{table}

\clearpage
\section{Molecular species: detections and non-detections}
\begin{figure*}[ht!]
    \centering
    \includegraphics[width=\textwidth]{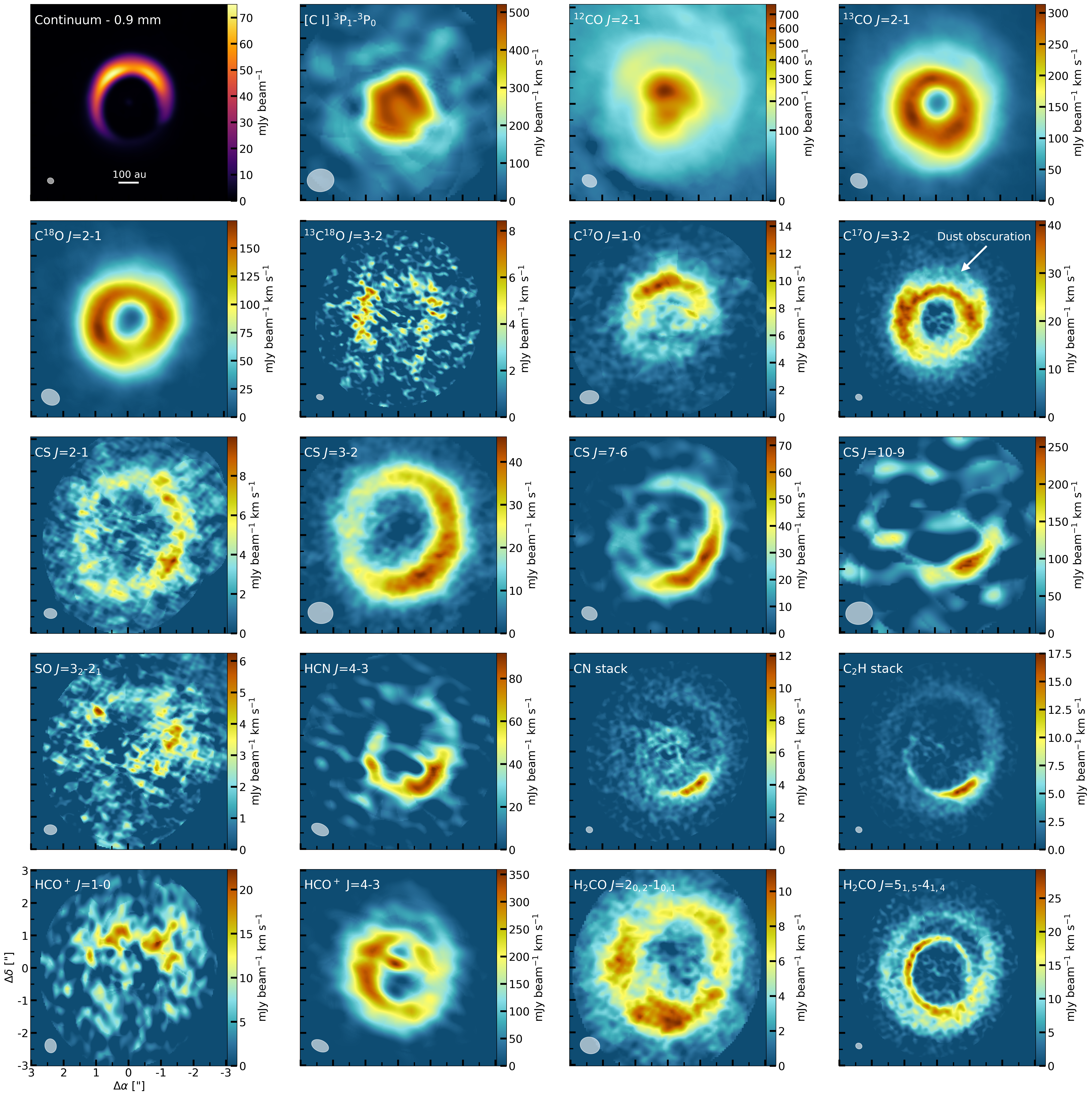}
    \caption{Integrated intensity maps of the dust continuum at 1.3~mm and the strongest detected molecular species in the disk of HD~142527. To increase the $S/N$-ratio of the respective integrated intensity maps, the images of the \ce{C^{17}O} $J$=1-0, \ce{C_2H} and \ce{CN} were created by stacking the detected transitions (two, four, and three transitions, respectively). The image of \ce{^{12}CO} $J$=2-1 was created with a different colour scaling to better highlight the weak extended emission. For the \ce{HCO^+} $J$=4-3 transition, we focused on imaging the Keplerian ring instead of the bright central spot. Subsequently, the integrated flux of the central spot may be lower than shown in \citet{TemminkEA23}. Additionally, the displayed images of \ce{C^{17}O} $J$=1-0, \ce{^{13}C^{18}O} $J$=3-2, \ce{CS} $J$=2-1 and $J$=3-2, and \ce{H_2CO} $J$=2$_{0,2}$-1$_{0,1}$ were created using a robust value of 2.0. The resolving beams are shown in the lower-left corner of each image.}
    \label{fig:Gallery}
\end{figure*}

\begin{table}[ht!]
    \centering
    \caption{Strongly detected molecular transitions in the disk of HD~142527.}
    \resizebox{\textwidth}{!}{%
    \begin{tabular}{c c c c c c c c c c}
        \hline\hline
        Molecule & Transition & Freq. & $\log\left(A_{ij}\right)$ & $E_\mathrm{up}$ & Robust & Beam                   & Peak flux & RMS                     & $\delta$V \\
                 &            & [GHz] & [s$^{-1}$]                 & [K]            &        & ["$\times$" (\degree)] & \multicolumn{2}{c}{[mJy~beam$^{-1}$]} & [km~s$^{-1}$] \\ 
        \hline
            \ce{C$^{17}$O} & $J$=1-0, $F$=7/2-5/2 & 112.358982 & -7.17 & 5.4 & 0.5 & 0.28$\times$0.24 (-85.9) & 11.1 & 1.8 & 0.30 \\
                           & $J$=1-0, $F$=5/2-5/2 & 112.360007 & -7.17 & 5.4 & 0.5 & 0.28$\times$0.24 (-85.9) & 13.2 & 1.8 & 0.30 \\
                           & $J$=3-2, $F$=9/2-7/2 & 337.060988 & -5.63 & 32.4 & 0.5 & 0.21$\times$0.18 (65.8) & 56.8 & 2.6 & 0.15 \\
            \ce{CS} & $J$=2-1 & 97.980953 & -4.78 & 7.1 & 0.5 & 0.25$\times$0.20 (77.2) & 10.0 & 1.7 & 0.30 \\
                    & - & - & - & - & 2.0 & 0.40$\times$0.31 (83.1) & 14.6 & 1.5 & 0.30 \\
                    & $J$=3-2 & 146.969029 & -4.22 & 14.1 & 0.5 & 0.39$\times$0.32 (88.0) & 32.1 & 3.0 & 0.30 \\
                    & - & - & - & - & 2.0 & 0.78$\times$0.66 (79.4) & 63.8 & 2.2 & 0.30 \\
            \ce{SO} & $J$=3$_2$-2$_1$ & 99.299870 & -4.95 & 9.2 & 2.0 & 0.40$\times$0.31 (85.8) & 9.1 & 1.5 & 0.30 \\
                    & - & - & - & - & 2.0 & 0.66$\times$0.47 (-85.4) & 2.4 & 0.4 & 2.00 \\
            \ce{CN} & $N$=3-2, $J$=5/2-3/2, $F$=7/2-5/2 & 340.031549 & -3.42 & 32.6 & 0.5 & 0.21$\times$0.18 (67.6) & 17.9 & 2.0 & 0.15 \\
                    & $N$=3-2, $J$=5/3-3/2, $F$=3/2-1/2 & 340.035408 & -3.49 & 32.6 & 0.5 & 0.21$\times$0.18 (67.6) & 20.9 & 2.0 & 0.15 \\
                    & $N$=3-2, $J$=7/2-5/2, $F$=7/2-5/2 & 340.247770 & -3.38 & 32.7 & 0.5 & 0.21$\times$0.18 (64.7) & 28.7 & 2.0 & 0.15 \\
            \ce{C_2H} & $N$=4-3, $J$=9/2-7/2, $F$=5-4 & 349.337707 & -3.88 & 41.9 & 0.5 & 0.20$\times$0.18 (64.5) & 49.3 & 2.2 & 0.15 \\
                      & $N$=4-3, $J$=9/2-7/2, $F$=4-3 & 349.339067 & -3.89 & 41.9 & 0.5 & 0.20$\times$0.18 (64.5) & 40.0 & 2.2 & 0.15 \\
                      & $N$=4-3, $J$=7/2-5/2, $F$=4-3 & 349.399374 & -3.90 & 41.9 & 0.5 & 0.20$\times$0.18 (64.5) & 42.9 & 2.2 & 0.15 \\
                      & $N$=4-3, $J$=7/2-5/2, $F$=3-2 & 349.400669 & -3.92 & 41.9 & 0.5 & 0.20$\times$0.18 (64.5) & 31.9 & 2.2 & 0.15 \\
            \ce{HCO^+} & $J$=1-0 & 89.188525 & -4.38 & 4.3 & 0.5 & 0.43$\times$0.35 (10.6) & 25.4 & 3.9 & 0.25 \\
            \ce{H_2CO} & $J$=2$_{0,2}$-1$_{0,1}$ & 145.602949 & -4.11 & 10.5 & 0.5 & 0.32$\times$0.27 (-85.3) & 9.5 & 1.2 & 0.30 \\
                       & - & - & - & - & 2.0 & 0.61$\times$0.51 (73.6) & 17.5 & 1.0 & 0.30 \\
                       & $J$=5$_{1,5}$-4$_{1,4}$ & 351.768645 & -2.92 & 62.5 & 0.5 & 0.20$\times$0.18 (63.4) & 19.6 & 0.9 & 1.00 \\
        \hline
    \end{tabular}%
    }
    \label{tab:Detections}
\end{table}

\subsection{Weak and/or tentative detections} \label{sec:WNDetections}
\begin{figure}[ht!]
    \centering
    \includegraphics[width=\textwidth]{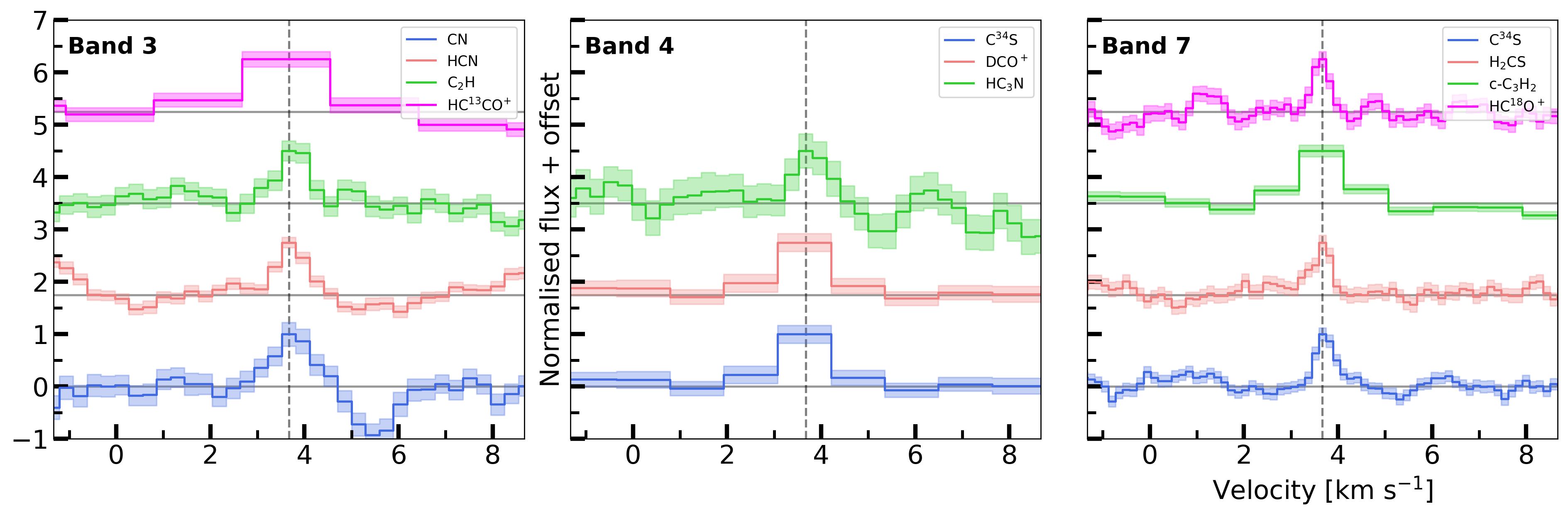}
    \caption{Normalised integrated spectra of the weak and tentative transitions detected with \textsc{GoFish}. The transitions found in Bands 3, 4, and 7 are, respectively, shown in the top, middle and bottom panels.}
    \label{fig:GoFish}
\end{figure}

\begin{figure*}[ht!]
    \centering
    \includegraphics[width=0.6\textwidth]{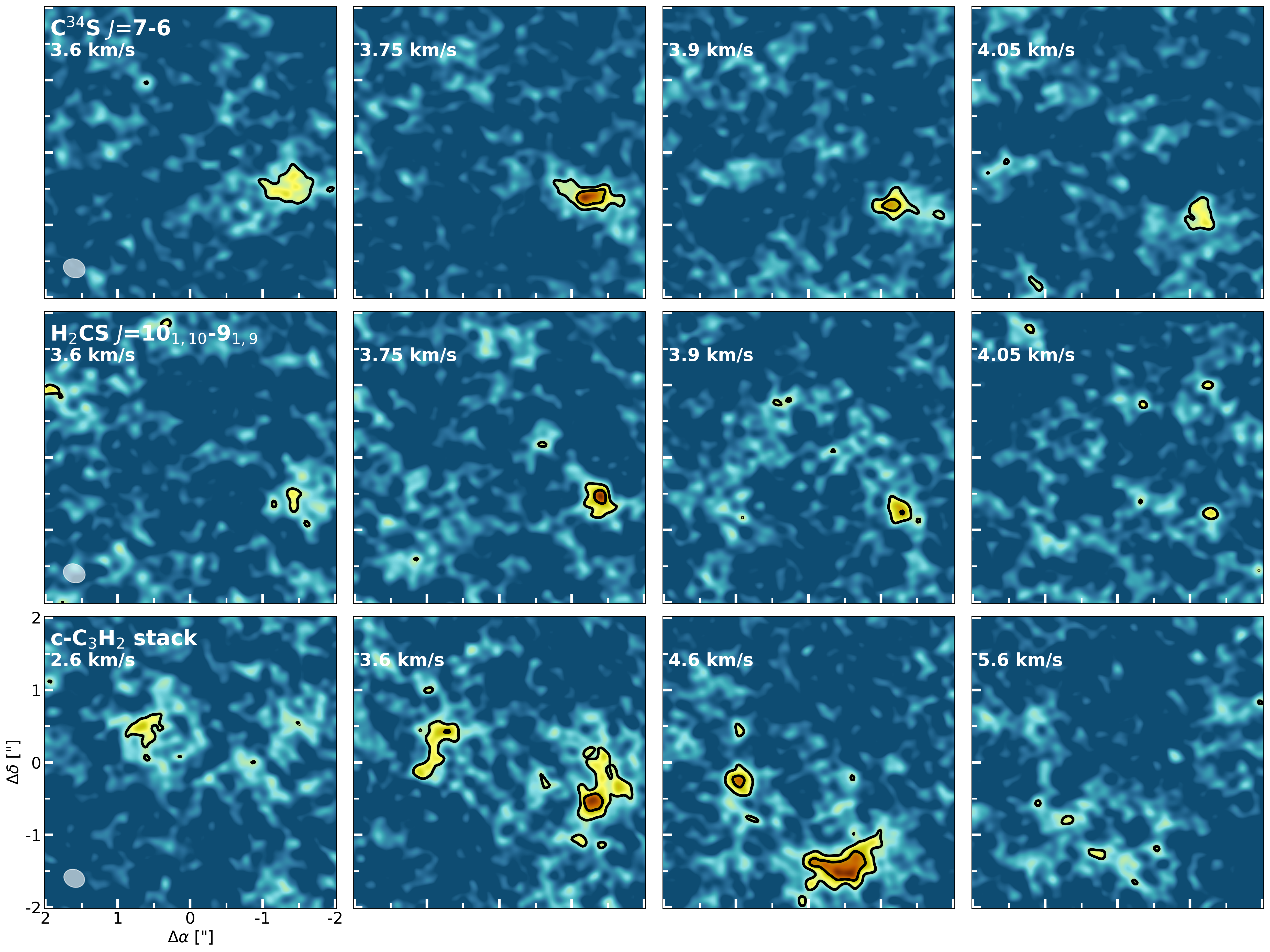}
    \caption{Selected channels of the weakly detected \ce{C^{34}S} $J$=7-6, \ce{H_2CS} $J$=10$_{1,10}$-9$_{1,9}$, and stacked \ce{c-C3H_2} transitions. The black contours indicate the 3$\sigma$ and 5$\sigma$ noise levels.}
    \label{fig:CM-WT}
\end{figure*}

With the aid of \textsc{GoFish} and subsequent visual inspection of the image cubes, we also report weak and/or tentative detections of \ce{C^{34}S} (two transitions), \ce{CN}, \ce{HCN}, \ce{C_2H}, \ce{H^{13}CO^+}, \ce{HC^{18}O^+}, \ce{DCO^+}, \ce{H_2CS}, \ce{c-C_3H_2}, and \ce{HC_3N}. The detections of \ce{CN}, \ce{HCN}, \ce{C_2H}, and \ce{c-C_3H_2} required the stacking of, respectively, three, three, three, and five separate transitions. The normalised integrated spectra obtained with \textsc{GoFish} are shown in Fig. \ref{fig:GoFish}, while the bottom part of Table \ref{tab:WeakDetections} contains information on the transitions. The additional peaks visible in the \textsc{GoFish} spectrum of \ce{HCN} (top panel in Fig. \ref{fig:GoFish}) are the result of stacking hyperfine transitions that lie closely together in frequency space. Finally, as the Band~7 transitions have stronger flux levels, Fig. \ref{fig:CM-WT} shows selected channel maps - in particular, channels containing the strongest flux levels - of the \ce{C^{34}S} $J$=7-6, \ce{H_2CS} $J$=10$_{1,10}$-9$_{1,9}$, and stacked \ce{c-C3H_2} transitions to reveal the emission morphology of these species. \\
\indent Following the discussion in Section \ref{sec:Molecules}, we highlight that the upper level energies ($E_\mathrm{up}\sim$77.2-96.5 K) and Einstein-A coefficients ($\log\left(A_\mathrm{ij}\right)$ ranges from -2.91 to -2.61) of the different \ce{c-C_3H_2} transitions are similar. Therefore, similar to the \ce{CN} and \ce{C_2H} emission, we expect the transitions to trace the same gas reservoirs. Even if the transitions would trace slightly different emitting layers, the disk's near face-on configuration ($i\sim$28.3\degree) will negate the majority of blurring effects due to potential different emitting heights. \\
\indent The integrated \textsc{GoFish} spectra of the \ce{C^{33}S} $J$=7-6 transition and the \ce{H_2 ^{13}CO} $J$=2$_{1,1}$-1$_{1,0}$ also hint at weak detections of these transitions. However, emission signatures could not be confirmed upon visual inspection of the image cubes. Therefore, we deem these two transitions to be non-detections.

\begin{table}[ht!]
    \centering
    \caption{Weakly or tentatively detected molecular transitions in the disk of HD~142527.}
    \resizebox{\textwidth}{!}{%
    \begin{tabular}{c c c c c c c c c c}
        \hline\hline
        Molecule & Transition & Freq. & $\log\left(A_{ij}\right)$ & $E_\mathrm{up}$ & Robust & Beam                   & Peak flux & RMS                     & $\delta$V \\
                 &            & [GHz] & [s$^{-1}$]                 & [K]            &        & ["$\times$" (\degree)] & \multicolumn{2}{c}{[mJy~beam$^{-1}$]} & [km~s$^{-1}$] \\ 
        \hline
            \ce{C^{34}S} & $J$=3-2 & 144.617101 & -4.24 & 13.9 & 2.0 & 0.63$\times$0.51 (68.6) & 2.6 & 0.5 & 0.30 \\
                         & $J$=7-6 & 337.396459 & -3.12 & 64.8 & 2.0 & 0.31$\times$0.25 (65.7) & 14.6 & 1.9 & 0.15 \\
            \ce{CN}$^{(\alpha)}$ & $N$=1-0, $J$=3/2-1/2, $F$=3/2-1/2 & 113.4881202 & -5.17 & 5.4 & 2.0 & 0.57$\times$0.41 (-86.2) & 4.4 & 1.0 & 0.30 \\
                                 & $N$=1-0, $J$=3/2-1/2, $F$=5/2-3/2 & 113.4909702 & -4.92 & 5.4 & 2.0 & - & - & - & 0.30 \\
                                 & $N$=1-0, $J$=3/2-1/2, $F$=1/2-1/2 & 113.4996443 & -4.97 & 5.4 & 2.0 & - & - & - & 0.30 \\
            \ce{HCN}$^{(\alpha)}$ & $J$=1-0, $F$=1-1 & 88.630416 & -4.62 & 4.3 & 2.0 & 0.45$\times$0.36 (81.4) & 4.4 & 0.8 & 0.30 \\
                     & $J$=1-0, $F$=2-1 & 88.631848 & -4.62 & 4.3 & 2.0 & - & - & - & 0.30 \\
                     & $J$=1-0, $F$=0-1 & 88.633936 & -4.62 & 4.3 & 2.0 & - & - & - & 0.30 \\
            \ce{C_2H}$^{(\alpha)}$ & $N$=1-0, $J$=3/2-1/2, $F$=1-1 & 87.284156 & -6.59 & 4.2 & 2.0 & 0.46$\times$0.36 (83.1) & 3.5 & 0.8 & 0.30 \\
                                   & $N$=1-0, $J$=3/2-1/2, $F$=2-1 & 87.316898 & -5.82 & 4.2 & 2.0 & - & - & - & 0.30 \\
                                   & $N$=1-0, $J$=3/2-1/2, $F$=1-0 & 87.328585 & -5.90 & 4.2 & 2.0 & - & - & - & 0.30 \\
            \ce{H^{13}CO^+} & $J$=1-0 & 86.7542884 & -4.41 & 4.2 & 2.0 & 0.47$\times$0.36 (82.6) & 2.2 & 0.5 & 0.30 \\
            \ce{HC^{18}O^+} & $J$=4-3 & 340.6306916 & -2.51 & 40.9 & 2.0 & 0.3$\times$0.26 (69.5) & 9.3 & 2.1 & 0.15 \\
            \ce{DCO^+} & $J$=2-1 & 144.077289 & -3.18 & 10.4 & 2.0 & 0.63$\times$0.51 (68.6) & 2.6 & 0.5 & 1.20 \\
            \ce{H_2CS} & $J$=10$_{1,10}$-9$_{1,9}$ & 338.083195 & -3.24 & 102.4 & 2.0 & 0.31$\times$0.25 (65.7) & 12.5 & 1.9 & 0.15 \\
            \ce{c-C_3H_2}$^{(\alpha)}$ & $J$=7$_{3,4}$-6$_{4,3}$ & 351.523317 & -2.91 & 77.2 & 2.0 & 0.30$\times$0.24 (64.9) & 2.6 & 0.4 & 1.00 \\
                                       & $J$=10$_{0,10}$-9$_{1,9}$ & 351.781578 & -2.61 & 96.5 & 2.0 & - & - & - & 1.00 \\
                                       & $J$=9$_{1,8}$-8$_{2,7}$ & 351.965969 & -2.67 & 93.3 & 2.0 & - & - & - & 1.00 \\
                                       & $J$=8$_{2,6}$-7$_{3,5}$ & 352.185542 & -2.76 & 86.9 & 2.0 & - & - & - & 1.00 \\
                                       & $J$=8$_{3,6}$-7$_{2,5}$ & 352.193656 & -2.76 & 86.9 & 2.0 & - & - & - & 1.00 \\
            \ce{HC_3N} & $J$=16-15 & 145.5609596 & -3.62 & 59.4 & 2.0 & 0.61$\times$0.51 (73.6) & 5.5 & 1.0 & 0.30 \\
        \hline
    \end{tabular}%
    }
    \label{tab:WeakDetections}
    \tablefoot{$^{(\alpha)}$: Only one value is listed for the beam, the peak flux, and the RMS for the transitions of \ce{CN}, \ce{HCN}, \ce{C_2H}, and \ce{c-C_3H_2} because we stacked the displayed transitions to ensure a detection.}
\end{table}

\subsection{Non-detections}
We also report non-detections of transitions of a wide variety of species: \ce{C^{33}S}, \ce{SO}, \ce{SiO}, \ce{CCS}, \ce{SO_2}, \ce{H^{13}CN}, \ce{HC^{15}N}, \ce{DCN}, \ce{H_2CO}, \ce{H_2 ^{13}CO}, \ce{H_2CS}, \ce{HC_3N}, \ce{CH_3OH}, and \ce{CH_3CN}. All information on the non-detected transitions, including the RMS of the observations, can be found in Table \ref{tab:NonDetections}. \\
\indent The most notable non-detection is that of \ce{CH_3OH}, even upon stacking 10 transitions. This is surprising since the molecule is now commonly observed in Herbig disks \citep{vdMarelEA21,BoothEA24a,BoothEA24b,BoothEA25,BoothEA26}. However, the stacked transitions have lower Einstein-A values compared to detected transitions \citep{BoothEA24b,BoothEA26,TemminkEA23}, but note that individual stronger transitions have not led to a detection (see, for example, \citealt{TemminkEA23}). Observations of \ce{CH_3OH} have often been linked to the sublimation of inherited molecular ices at cavity walls, as the disks are generally too warm for in-situ formation (see, for example, \citealt{vdMarelEA21,BoothEA25,BoothEA26}). Since \ce{CH_3OH} does not have efficient gas-phase formation routes, the hydrogenation of frozen-out \ce{CO} molecules is considered to be the most common formation pathway \citep{WK02,FuchsEA09,SantosEA22}. The RADMC-3D model of \citet{TemminkEA23} suggests that \ce{CO} can only be frozen out at the outer edge of the dust trap. Therefore, the in-situ formation of \ce{CH_3OH} in the disk of HD 142527 is unlikely, and any \ce{CH_3OH} present in the disk should also be the result of inheritance. Our non-detection of \ce{CH_3OH}, suggests that the dust trap is located too far from the host star to be significantly heated for \ce{CH_3OH}-ice to sublimate. Additionally, the vertical turbulence may be too weak for the icy dust grains to be lifted to the higher elevated layers where sublimation may occur. 
\begin{table}[ht!]
    \centering
    \caption{Non-detected molecular transitions in the disk of HD~142527.}
    \resizebox{\textwidth}{!}{%
    \begin{tabular}{c c c c c c c c c}
        \hline\hline
        Molecule & Transition & Freq. & $\log\left(A_{ij}\right)$ & $E_\mathrm{up}$ & Robust & Beam                   & RMS               & $\delta$V \\
                 &            & [GHz] & [s$^{-1}$]                 & [K]            &        & ["$\times$" (\degree)] & [mJy~beam$^{-1}$] & [km~s$^{-1}$] \\ 
        \hline
        \ce{C^{33}S} & $J$=7-6 & 340.0525755 & -3.10765 & 65.3 & 2.0 & 0.31$\times$0.26 (68.8) & 1.8 & 0.15 \\
        \ce{SO} & $J$=2$_2$-1$_1$ & 86.093950 & -5.27985 & 19.3 & 2.0 & 0.47$\times$0.37 (82.5) & 0.5 & 2.0 \\
        \ce{SiO} & $J$=2-1 & 86.846985 & -4.5334 & 6.3 & 2.0 & 0.47$\times$0.36 (82.6) & 0.5 & 2.0 \\
        \ce{CCS}$^{\alpha}$ & $N$=7-6, $J$=6-5 & 86.181391 & -4.55632 & 23.3 & 2.0 & 0.47$\times$0.37 (82.6) & 0.5 & 2.0 \\
                            & $N$=8-7, $J$=7-6 & 99.866521 & -4.3562 & 28.1 & 2.0 & 0.65$\times$0.46 (-85.8) & 0.4 & 2.0 \\ 
                            & $N$=9-8, $J$=8-7 & 113.410186 & -4.18482 & 33.6 & 2.0 & - & - & 2.0 \\ 
        \ce{SO_2}$^{\alpha}$ & $J$=10$_{6,4}$-11$_{5,7}$ & 350.862756 & -4.35623 & 138.8 & 2.0 & 0.30$\times$0.24 (64.9) & 0.5 & 1.0 \\
                             & $J$=5$_{3,3}$-4$_{2,2}$ & 351.2572233 & -3.47398 & 35.9 & 2.0 & - & - & 1.0 \\
                             & $J$=14$_{4,10}$-14$_{3,11}$ & 351.8738732 & -3.46440 & 135.9 & 2.0 & - & - & 1.0 \\
        \ce{H^{13}CN} & $J$=1-0 & 86.3399214 & -4.65260 & 4.1 & 2.0 & 0.47$\times$0.37 (82.6) & 0.5 & 2.0 \\
        \ce{HC^{15}N} & $J$=1-0 & 86.0549664 & -4.65693 & 4.1 & 2.0 & 0.47$\times$0.37 (82.5) & 0.5 & 2.0 \\
        \ce{DCN} & $J$=2-1 & 144.8280015 & -3.89786 & 10.4 & 2.0 & 0.63$\times$0.51 (68.8) & 0.5 & 1.2 \\
        \ce{H_2CO} & $J$=6$_{1,5}$-6$_{1,6}$ & 101.332991 & -5.80378 & 87.6 & 2.0 & 0.39$\times$0.30 (84.9) & 1.6 & 0.3 \\
                   & -                       & -          & -        & -    & 2.0 & 0.67$\times$0.47 (-85.5) & 0.4 & 2.0 \\
        \ce{H_2^{13}CO} & $J$=2$_{1,1}$-1$_{1,0}$ & 146.6356717 & -4.22279 & 22.4 & 2.0 & 0.61$\times$0.51 (74.5) & 1.1 & 0.3 \\
        \ce{H_2CS} & $J$=3$_{1,3}$-2$_{1,2}$ & 101.4778095 & -4.89960 & 22.9 & 2.0 & 0.39$\times$0.30 (85.0) & 1.6 & 0.3 \\
        \ce{HC_3N} & $J$=11-10 & 100.076392 & -4.10963 & 28.8 & 2.0 & 0.65$\times$0.46 (-85.6) & 0.4 & 2.0 \\
        \ce{CH_3OH}$^{\alpha}$ & $J$=3$_{-0,3}$-2$_{-0,2}$ & 145.093754 & -4.91 & 27.1 & 2.0 & 0.59$\times$0.49 (71.3) & 0.5 & 1.2 \\
                               & $J$=3$_{0,3}$-2$_{0,2}$ & 145.103185 & -4.91 & 13.9 & 2.0 & - & - & 1.2 \\
                               & $J$=3$_{2,2}$-2$_{2,1}$ & 145.124332 & -5.16 & 51.6 & 2.0 & - & - & 1.2 \\
                               & $J$=3$_{-1,2}$-2$_{-1,1}$ & 145.131864 & -4.95 & 35.0 & 2.0 & - & - & 1.2 \\
                               & $J$=6$_{2,4}$-7$_{1,7}$ & 156.127544 & -5.18 & 86.5 & 2.0 & - & - & 1.2 \\
                               & $J$=8$_{-0,8}$-8$_{1,8}$ & 156.488902 & -4.75 & 96.6 & 2.0 & - & - & 1.2 \\
                               & $J$=7$_{-0,7}$-7$_{1,7}$ & 156.828517 & -4.73 & 78.1 & 2.0 & - & - & 1.2 \\
                               & $J$=5$_{-0,5}$-5$_{1,5}$ & 157.178987 & -4.69 & 47.9 & 2.0 & - & - & 1.2 \\
                               & $J$=4$_{-0,4}$-4$_{1.4}$ & 157.246062 & -4.68 & 36.3 & 2.0 & - & - & 1.2 \\
                               & $J$=1$_{-0,1}$-1$_{1,1}$ & 157.270832 & -4.66 & 15.4 & 2.0 & - & - & 1.2 \\
                               & $J$=4$_{-0,4}$-3$_{1,3}$ & 350.687662 & -4.06 & 36.3 & 2.0 & 0.30$\times$0.24 (64.9) & 0.5 & 1.0 \\
                               & $J$=1$_{1,1}$-0$_{0,0}$ & 350.905100 & -3.48 & 16.8 & 2.0 & - & - & 1.0 \\
        \ce{CH_3CN}$^{\alpha}$ & $J$=19$_{3}$-18$_{-3}$ & 349.3932973 & -2.44 & 232.0 & 2.0 & 0.30$\times$0.24 (65.6) & 0.5 & 1.0 \\
                               & $J$=19$_{2}$-18$_{2}$ & 349.4268499 & -2.43 & 196.3 & 2.0 & - & - & 1.0 \\
                               & $J$=19$_{1}$-18$_{1}$ & 349.4469869 & -2.43 & 174.9 & 2.0 & - & - & 1.0 \\
                               & $J$=19$_{0}$-18$_{0}$ & 349.4537001 & -2.43 & 167.7 & 2.0 & - & - & 1.0 \\
        \hline
    \end{tabular}%
    }
    \tablefoot{$^{(\alpha)}$: In an attempt to detect the molecular emission, we stacked the various transitions of \ce{CCS}, \ce{SO_2}, and \ce{CH_3OH}. Therefore, only one value is listed for the beam and the RMS.}
    \label{tab:NonDetections}
\end{table}

\section{Dust brightness temperature and spectral index} \label{sec:Dust-OD}
\begin{figure}[ht!]
	\centering
	\includegraphics[width=0.7\textwidth]{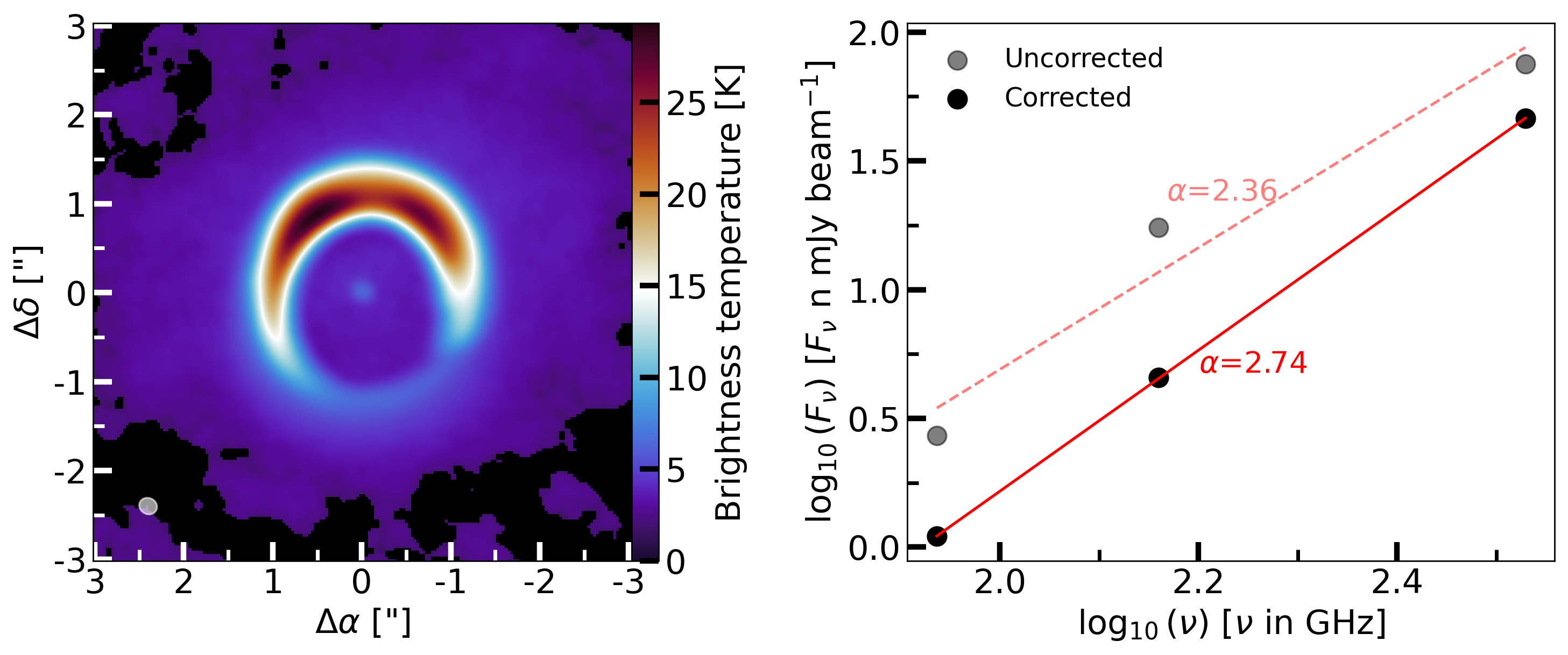}
	\caption{Brightness temperature of the dust continuum at 0.9~mm is shown in the left panel. The resolving beam is shown in the bottom left corner. The right panel displays the fit for the spectral index for the peak fluxes (gray datapoints) and the corrected peak fluxes (black datapoints).}
	\label{fig:Dust-OD}
\end{figure}

The left panel of Figure \ref{fig:Dust-OD} displays the brightness temperature of the Band~7 (0.9~mm) continuum emission. The brightness temperature peaks at $T_\mathrm{b,peak}\sim$29.3~K, which suggests, in the case of optically emission, that the dust trap is slightly warmer than the \ce{CO} sublimation temperature. The right panel of Figure \ref{fig:Dust-OD} shows the peak fluxes of the Band~3, Band~4, and Band~7 continua (gray datapoints), and those fluxes corrected to the same circular beam with a major axis of 0.15" (black datapoints). The spectral index ($\alpha$) has been derived by fitting a straight line through these datapoints:
\begin{align}
	\log_{10}\left(F_\nu\right) = \alpha\log_{10}\left(\nu\right) + C.
\end{align}
Here, $\nu$ denotes the frequency in GHz and $C$ is a constant that sets the intercept. Table \ref{tab:SpectralIndex} lists the used peak fluxes and frequencies, and retrieved values for the fitted parameters ($\alpha$ and $C$).

\begin{table}[ht!]
	\centering
	\caption{Peak fluxes for each continuum image and fitted values for the spectral index.}
	\begin{tabular}{c | c c c | c c }
		\hline\hline
		& $F_{86\mathrm{GHz}}$ & $F_{144\mathrm{GHz}}$ & $F_{338\mathrm{GHz}}$  & $\alpha$ & $C$ \\
		& \multicolumn{3}{c|}{[mJy~beam$^{-1}$]} & & \\
		\hline
		Uncorrected & 2.7 & 17.4 & 75.1 & 2.36 & -4.0 \\
		Corrected   & 1.1 & 4.5 & 46.2 & 2.74 & -5.26 \\
		\hline
	\end{tabular}
	\label{tab:SpectralIndex}
\end{table}

\section{\ce{H_2CO} channel maps}
\begin{figure*}[ht!]
    \centering
    \includegraphics[width=\textwidth]{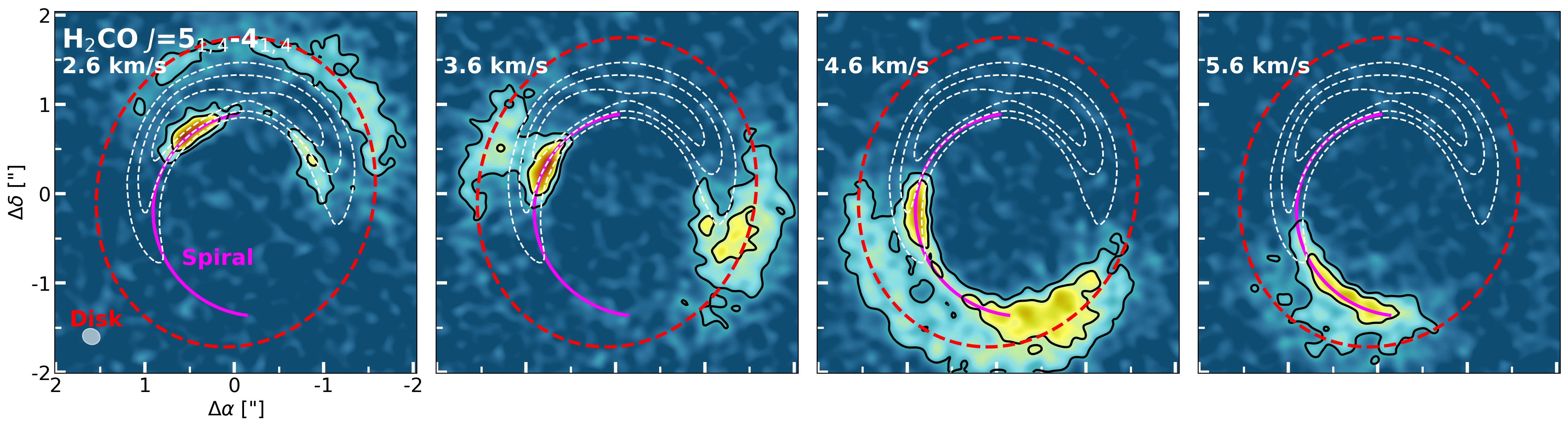}
    \caption{Selected channels of the \ce{H_2CO} $J$=5$_{1,5}$-4$_{1,4}$ transition. We have labeled the different contributions from the spiral (pink) and the disk (red). The black contours indicate the 5$\sigma$ and 10$\sigma$ noise levels, while the white, dashed contours indicate the continuum emission at flux levels of 25\%, 50\%, and 75\%.}
    \label{fig:CM-H2CO}
\end{figure*}

The \ce{H_2CO} channel maps (Fig. \ref{fig:CM-H2CO}) reveal that the observed spiral-like feature is real and not the cause of continuum oversubtraction effects. In particular, the channel maps at 4.6~km~s$^{-1}$ and 5.6~km~s$^{-1}$ display that the emission in the southern region of the disk moves radially inwards while there is no strong continuum emission present. Therefore, we consider the spiral-like feature to be real, yet we cannot rule out that continuum oversubtraction effects are affecting the emission and that, as a result, the feature appears more prominently.

\section{Tracing the \ce{^{12}CO} spiral arms} \label{sec:Spirals}
\begin{figure*}[ht!]
    \centering
    \includegraphics[width=\textwidth]{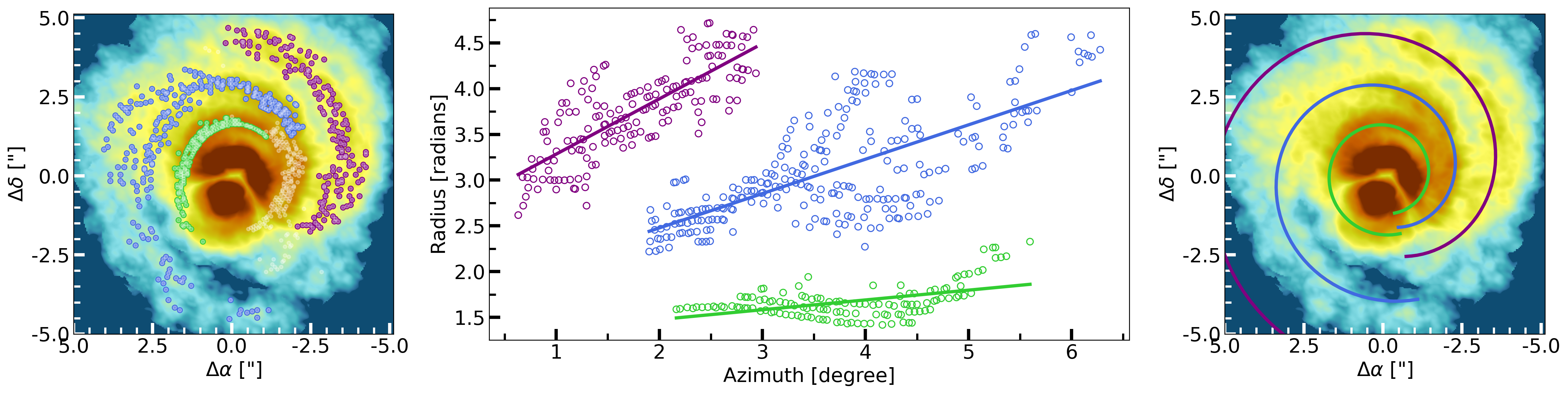}
    \caption{Left: Peak intensity map of the \ce{^{12}CO} $J$=2-1 transition with the identified peaks in the residual brightness channel maps (left panel). The coloured data points (blue, green, or purple) are the data points we have assigned to a spiral feature. Middle: Archimedean linear spiral fits  (solid lines) to the identified spirals. Right: Fitted spiral plotted on top of the \ce{^{12}CO} peak intensity map.}
    \label{fig:12CO-Spirals}
\end{figure*}

\noindent To trace the spiral arms seen in the \ce{^{12}CO} $J$=2-1 line emission, we combine previously used techniques \citep{TeagueEA19,GargEA21,WoelferEA21}. In particular, we investigate the spiral arms in the brightness temperature channel maps after subtracting off the averaged brightness temperature map in each channel. Using azimuthal increments of 3\degree, we subsequently identify radial peaks whose flux exceeds at least 3$\times$ the channel RMS and whose residual brightness temperature exceeds at least half the median flux of the averaged bright temperature map. Furthermore, we require the peaks to be spaced at least 10 pixels apart. Through visual inspection of the channels, we assign the identified peaks to a spiral or leave them unassigned. The left panel of Fig. \ref{fig:12CO-Spirals} shows the peak intensity map of the \ce{^{12}CO} emission, where the coloured dots (blue, green, or purple) are assigned to a spiral (three in total) and the white ones have been left unassigned. We note that a set of identified peaks, potentially connected to the blue and/or green spirals, has been left unassigned, due to the potential conjunction of these spirals, making it nearly impossible to assign the peaks to either one of the spirals. To avoid assigning data points to the wrong spiral, we have left these data points unassigned. \\
\indent To further investigate the spirals, we converted the Cartesian (pixel) coordinates of the peaks to cylindrical coordinates ($r$, $\phi$), not taking the disk's inclination and position angle into account, and we have fitted an Archimedean linear spiral to these coordinates:
\begin{align}
    r = a + b\phi.
\end{align}
We note that the cylindrical coordinates were rotated by 105$\degree$ to ensure that the spirals were not broken up along the $\phi$-coordinate. The resulting fits are shown in the middle panel of Fig. \ref{fig:12CO-Spirals}, and the fitted spirals, converted back to the Cartesian coordinates, are also shown on top of the \ce{^{12}CO} in the right panel of Fig. \ref{fig:12CO-Spirals}. Furthermore, the fit values for $a$ and $b$ for each of the three spirals are listed in Table \ref{tab:SpiralFits}. \\
\begin{table}[ht!]
    \centering
    \begin{tabular}{c c c}
        \hline\hline
        Spiral & $a$ ["] & $b$ [" rad$^{-1}$] \\
        \hline
        Blue & 1.73 & 0.38 \\
        Purple & 2.68 & 0.60 \\
        Green & 1.26 & 0.11 \\
        \hline
    \end{tabular}
    \caption{Best-fitting parameters for the Archimedean linear spirals.}
    \label{tab:SpiralFits}
\end{table}

\indent Our identified spirals match those of \citet{GargEA21}. In particular, our blue spiral is a combination of their spirals 1 and 3 (S1 and S3), of which they noted that they could be connected. Our purple spiral matches their spiral 2 (S2), while our green spiral matches their spiral 4 (S4). Furthermore, \citet{GargEA21} discussed that their S1 and S4 (or our blue and green) may also be connected. This is also hinted at by our traced spirals, but this cannot be confirmed due to the molecular emission being absorbed by the foreground cloud. \\
\begin{figure}[ht!]
    \centering
    \includegraphics[width=0.5\textwidth]{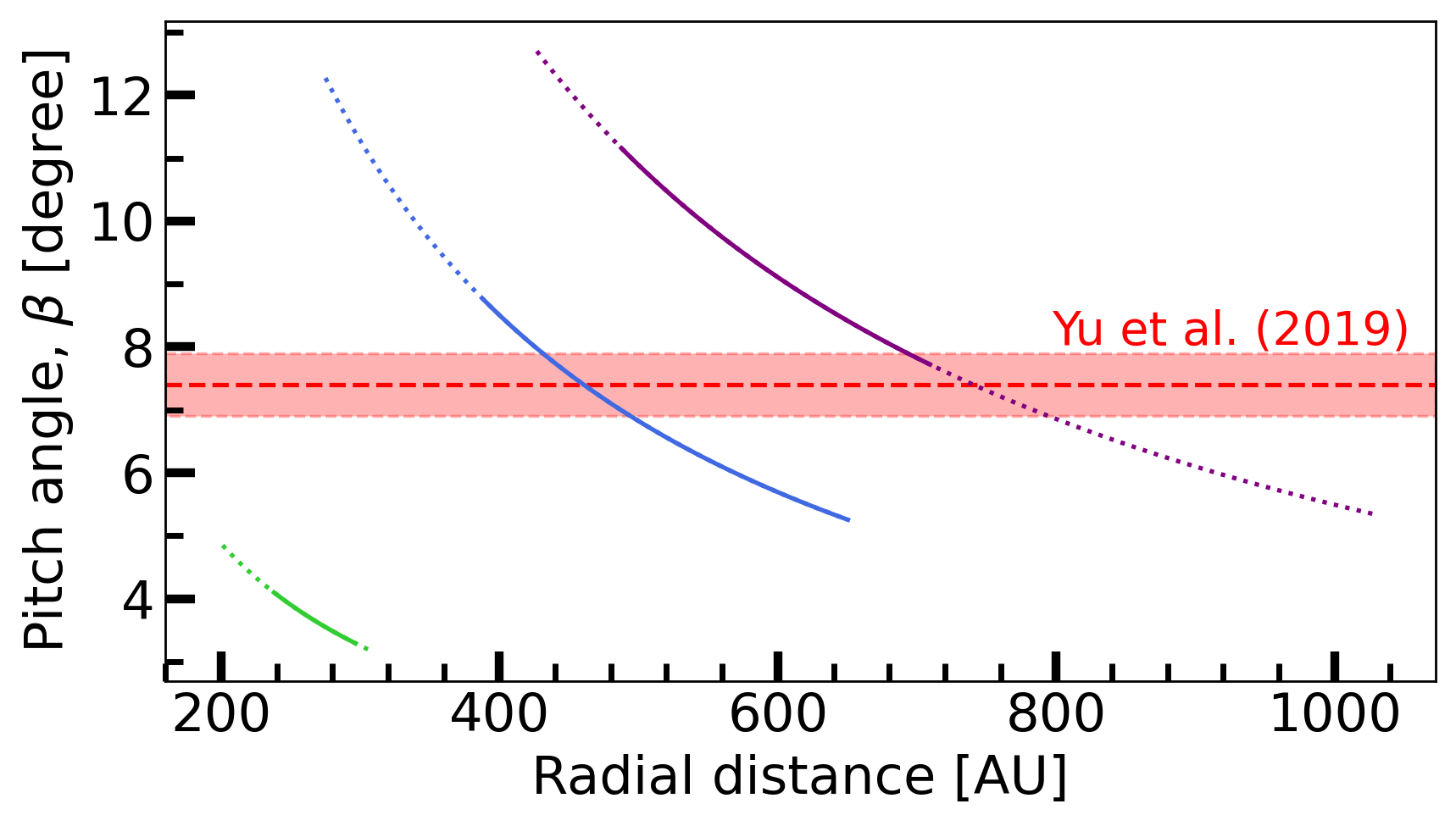}
    \caption{Pitch angles of the identified spiral arms. The horizontal red line indicates the value found by \citet{YuEA19} for the spiral arm observed in the scattered light.}
    \label{fig:Spirals-PitchAngles}
\end{figure}

\indent Following the procedure given by \citet{WoelferEA21}, we also calculate the pitch angle ($\beta$) for the linear fits. The pitch angle is given by the following formula:
\begin{align}
    \tan\left(\beta\right) = \left|\frac{\mathrm{d}r}{\mathrm{d}\phi}\right|\cdot\frac{1}{r}.
\end{align}
The pitch angles are calculated for all three identified spirals, and the results are shown in Fig. \ref{fig:Spirals-PitchAngles}. As a comparison, \citet{YuEA19} derived a pitch angle of $\beta$=7.4$\pm$0.5 for the spiral arm observed in the scattered light image. This matches fairly well with the pitch angles derived for our spirals.

\section{Deprojected maps: \ce{C_2H} and \ce{H_2CO}}
\begin{figure*}[ht!]
    \centering
    \includegraphics[width=0.85\textwidth]{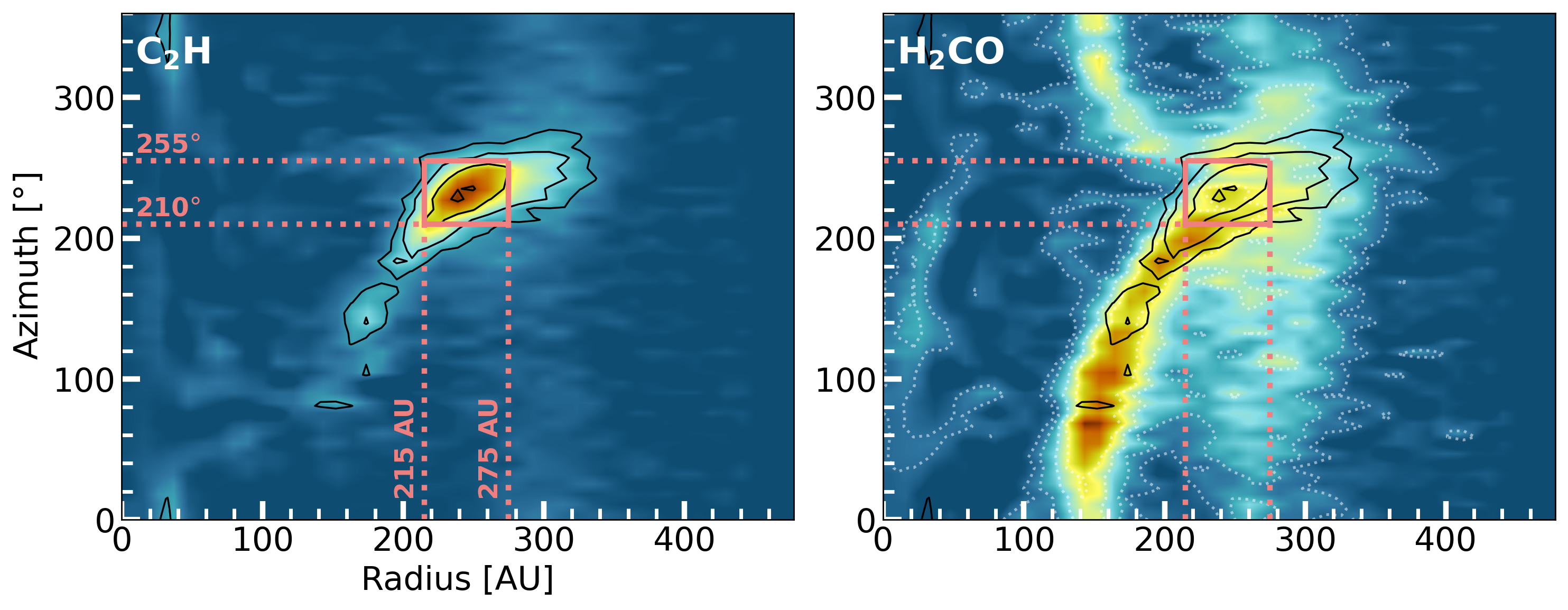}
    \caption{Deprojected versions of the integrated intensity maps of the stacked \ce{C_2H} and \ce{H_2CO} $J$=5$_{1,5}$-4$_{1,4}$ transitions. The pink lines indicate the visually approximated extent of the \ce{C_2H} peak emission. Furthermore, the solid black contours indicate the 3$\sigma$, 5$\sigma$, 10$\sigma$, and 15$\sigma$ \ce{C_2H} emission, whereas the white dotted contours represent those of the \ce{H_2CO} emission.}
    \label{fig:Deprojected}
\end{figure*}

\end{appendix}

\end{document}